\documentclass[preprint,review,14pt]{elsarticle}
\usepackage[utf8x]{inputenc}
\usepackage{lineno}
\usepackage{graphicx}
\usepackage{booktabs}
\usepackage{epsfig}
\usepackage{amsmath,amsfonts, amsthm, mathrsfs, amssymb, bbm}
\usepackage{times}
\usepackage{apalike}
\setcitestyle{authoryear,open={(},close={)}}
\usepackage{subfigure}
\journal{arXiv preprint}
\usepackage[table,xcdraw]{xcolor}
\usepackage{flafter}

\begin{document}

\begin{frontmatter}

\title{Evolutionary Dynamics of Gig Economy Labor Strategies under Technology, Policy and Market Influence}

\author[DartMath]{Kevin Hu}
\ead{kevn.hu@gmail.com}
\author[DartMath,DartBMDS]{Feng Fu\corref{ff}}
\ead{fufeng@gmail.com}

\address[DartMath]{Department of Mathematics, Dartmouth College, Hanover, NH 03755, USA}

\address[DartBMDS]{Department of Biomedical Data Science, Geisel School of Medicine at Dartmouth, Lebanon, NH 03756, USA}

\cortext[ff]{Corresponding author at: 27 N. Main Street, 6188 Kemeny Hall, Department of Mathematics, Dartmouth College, Hanover, NH 03755, USA. Tel: +1 (603) 646 1047, Fax: +1 (603) 646 1312}

\begin{abstract}
The emergence of the modern gig economy introduces a new set of employment considerations for firms and laborers that include various trade-offs. With a game-theoretical approach, we examine the influences of technology, policy and markets on firm and worker preferences for gig labor. Theoretically, we present a new extension to the replicator equation and model oscillating dynamics in two-player asymmetric bi-matrix games with time-evolving environments, introducing concepts of the attractor arc, trapping zone and escape. We demonstrate how changing market conditions result in distinct evolutionary patterns for gig-labor preferences across high and low skill work-forces, which we explain through their differing sensitivities to market-driven consumer demand and financial incentives among other considerations. Informing tensions regarding the future of this new employment category, we present a novel payoff framework to analyze the role of technology on the growth of the gig economy. Finally, we explore regulatory implications within the gig economy, demonstrating how intervals of lenient and strict policy alter firm and worker sensitivities between gig and employee labor strategies. 

\end{abstract}

\begin{keyword}
 Evolutionary game theory \sep Gig worker  \sep Evolutionary economics  \sep Social learning  \sep Oscillatory dynamics \end{keyword}

\end{frontmatter}


\section{Introduction}
\label{intro}

With economic prevalence that extends to the labor markets of the early Roman Empire~\citep{temin2004labor}, the concept of contract work has existed for millennia, manifesting in different forms across societies and temporal interludes~\citep{applebaum1992concept}. In recent decades, contract or 'gig' work has emerged as a commanding employment category in the United States, having captured more than one third of the labor market by 2018~\citep{gallup2018gig}. At the cornerstone of this development are online labor marketplaces that facilitate the exchange of talent and capital between firms and workers, effectively decreasing hiring frictions and increasing labor liquidity~\citep{benjaafar2020operations,donovan2016does,kuhn2016rise}. The result of this infrastructure furtherance takes form in a novel, complementary contract-based labor market monikered the sharing, collaborative or gig economy~\citep{burtch2018can, sundararajan2014peer, ravenelle2017sharing, yaraghi2017current}.
	
Specific to the rise of gig work in the digital era, the modern gig economy enables firms to digitally outsource tasks and processes to remote workforces and match independent skill sets to specific labor needs~\citep{aloisi2015commoditized}. In addition to specific or high-skill labor, the gig economy capacitates firms in employing legions of commodity-skill gig laborers to operate their service offerings. For instance, ride-share companies such as Lyft and Uber leverage contractual gig drivers in their businesses, ultimately re-engineering cheaper, on-demand product offerings~\citep{amey2011real,janasz2017future}. There is, however, a trade-off; the commitment to cheaper pricing with gig operators may come at the expense of product and service quality~\citep{prassl2018humans}. On the labor supply side, autonomy, self governance and overall increased flexibility form the gravitational kernel that captivates new workers and persuades them to participate in the gig economy~\citep{lehdonvirta2018flexibility, broughton2018experiences}. However, gig workers lack the income stability and labor protections such as union rights and insurance benefits conferred with employee status~\citep{oranburg2018unbundling}. Thus, there exists a vast multiplex of considerations for firms and laborers regarding their labor decisions.

Beyond the firm and individual, there are several macro factors at play. The dynamics of firm and worker labor preferences are saddled at the nexus between \emph{market conditions, technology and policy}. Noting select events of American economic history from the last century, we see a pattern wherein which the importance and popularity of contract work fluctuate as a result of several economic factors. Notably, during the post Great Depression and World War II period, workers sought out an auxiliary arrangement, a reconstitution of work and enterprise, in a pursuit of autonomy and stability~\citep{hyman2018temp}. Over the last century, this campaign for autonomy, not contemporary digital applications, set the foundation for the modern gig economy~\citep{hyman2018temp}. More recently, the Great Recession in 2008 resulted in a shift in consumerism manifesting in decreased consumption of services, demonstrating the impact that economic downturns have on consumer behavior~\citep{giroud2017firm}. Indeed, decreased consumer consumption directly affects demand for commodity skills and contractual labor. In particular, commodity or low skill workers are most adversely affected in bear market conditions~\citep{levine2009labor}. Further, present-day gig workers recognize that the structural forces of economic recessions restrict their autonomy; when demand for work declines, gig laborers remain persistently available to compete for limited contracts, thereby disqualifying any scheduling flexibility~\citep{lehdonvirta2018flexibility}. A market-labor pattern emerges across history and informs us how evolving market cycles shape the labor landscape. We aim to apply historical observations and existing literature to deeply explore market influences on gig economy labor strategies.
	
There is also compelling evidence to believe that technological advancements may engender the future growth or stagnation of the gig economy. On the one hand, there is an expectation that the gig economy will continue to grow with the introduction of new sharing platforms and businesses~\citep{yaraghi2017current, manyika2016independent}. Additionally, there is contention that frontier technologies such as blockchain will accelerate the de-centralization of enterprise economies~\citep{pouwelse2017laws}, further enabling the growth of the gig economy~\citep{ kursh2016adding}. On the other hand, there exists a growing accord in scholarship that artificial intelligence (AI) will displace many human operators~\citep{benjaafar2020operations}, especially those with commodity skills~\citep{acemoglu2018artificial}. The rapid acceleration of AI may implicate the displacement of gig workers, for instance, the substitution of ride-sharing drivers with the introduction of autonomous vehicles~\citep{benjaafar2020operations}. A question remains as to whether these displaced workers will reenter the workforce as employees or gig workers. Seemingly, the influence of technology on the future of the gig economy depends on a constellation of co-developing technologies racing to fruition.

In recent years, there has been growing effort in studying the gig economy, which provides useful insights that address labor preferences, policy design, the role of technology and wide-ranging socioeconomic implications. 

Among others, one main approach used to study the gig economy is ethnography with various statistical methods. Much has been explored regarding influences on firm and worker gig-economy incentives. Allon \textit{et al.} collaborate with a ride-sharing platform to investigate behavioral and economic incentives for gig workers, noting a prioritization of an earnings goal over the number of hours worked and a willingness to work more with more hours worked \citep{allon2018impact}. Lehdonvirta explores flexibility in the gig economy, reiterating emphasis on the income-target and finds support that worker autonomy depends on a large availability of work \citep{lehdonvirta2018flexibility}. Burtch \textit{et al.} study how gig-economy platforms influence entrepreneurial activity, finding that gig platforms reduce total entrepreneurial activity as these platforms provide prospective entrepreneurs an additional stream of income. Leung examines hiring in the gig economy as a learning experience, noting that firms expressed loss-aversion behaviors when responding to positive and negative hiring experiences \citep{leung2018learning}. Exploring hiring across the global gig-economy, Galperin \textit{et al.} note discriminatory geographical preferences in firms' hiring preferences \citep{galperin2017geographical}.

A number of works have also explored the role of high-skill contractors. Anderson and Bidwell investigate managerial roles in the gig economy, exploring a friction in the cohesion of managerial responsibilities and contract work arrangements \citep{anderson2019outside}; they find that managerial contractors experience more flexibility but reduced pay. In contrast, other studies suggest that high skill contractors earn higher salaries than employees \citep{houseman2003role, pearce1993toward}. Barley and Kunda find that high skill contractors in technical professions earn more than regular employees \citep{barley2006gurus}, and Bidwell and Bricoe find that technical contractors working in Internet Technology (IT) earn the same as employees \citep{bidwell2009contracts}.

Academic research on the gig economy has also extensively embraced concerns in policy, technology and economics. Friedman argues that the growth of the gig economy requires new social policy as economic risks are shifted from the firm to the laborer~\citep{friedman2014workers}. Todoli-Signes examines the gig worker's need for protection and details regulatory concern around working hours, minimum wage, child labor bans and annual leave among other areas of apprehension~\citep{todoli2017end}. Stewart and Stanford investigate five regulatory mechanisms in the gig economy such as the creation of a new independent worker category or the provision of workers' rights, reviewing the pros and cons of each framework~\citep{stewart2017regulating}. While research focusing on regulation and policy collectively exhibit a concern regarding the gig economy, many scholarly works on technological developments concentrate on drivers of growth for this new employment sector. In this work, we consolidate many of the aforementioned areas of research and, from game theoretical perspective, study the influence of policy, technology and market changes on firm and laborer preferences in the gig economy.


While classical game theory developed to address questions in economics~\citep{nash1950bargaining, von2007theory}, the field of evolutionary game theory, a theoretical extension that models how populations change strategies over time~\citep{cressman2014replicator}, finds its roots in biology~\citep{smith1973logic, cressman2014replicator}. Since its inception in 1973~\citep{smith1973logic}, evolutionary game theory has broadened in application beyond its early biological origins to study social interactions and population behaviors across various academic fields~\citep{cressman2014replicator, alexander2002evolutionary,bear2016intuition,rand2011evolution,apicella2019evolution,perc2017statistical}. 

In evolutionary dynamics, the approach of replicator equation is most notable. Originally presented by Taylor and Jonker in 1978~\citep{taylor1978evolutionary} and formally named by Schuster and Sigmund \citep{schuster1983replicator}, the replicator equation determines the evolution of the composition of strategies in a population~\citep{traulsen2009stochastic}. 


As the modern gig economy grows out of its unhampered infancy, policy makers and researchers alike are presented the question of how this market should be regulated~\citep{aloisi2015commoditized, de2016introduction}. Undefined ordinance allows new competitors leveraging gig work to play by different rules than industry incumbents, a result of ambiguous labor laws that enable firms to shift economic burdens onto the gig laborer \citep{todoli2017end, johnston2018organizing, isaac2014disruptive}. In industry, some governments have mandated that firms more closely classify gig workers as employees, a decree that demands additional securities for gig laborers \citep{dubal2017winning, semuels2018happens}. The question as to whether or how this new labor sector should be policed remains unanswered, an inquiry of apprehension we aim to inform in the present work.
     
Using a game-theoretical approach, we investigate both firm and individual labor considerations as well as the economic influences of markets, technology and policy on labor preferences in the gig economy. In this paper, we present a new extension to the replicator equation, oscillating replicator dynamics with attractor arcs, formally, an oscillating replicator dynamics of  two player asymmetric bi-matrix games with time-evolving environment. Previous studies have analyzed oscillating tragedy of the commons for evolutionary games with environmental feedback~\citep{wu2011moving,weitz2016oscillating,hauert2019asymmetric,shao2019evolutionary,tilman2020evolutionary,wang2020steering}. Using the framework of replicator dynamics \citep{traulsen2009stochastic}, we model the evolutionary behavior of firm and laborer preferences for gig strategies. Incorporating assumptions founded on existing scholarship, we generate payoff matrices that reflect the incentives of each labor strategy (i.e., hiring a gig worker or employee) given a specific market condition. While we base our model on existing works in evolutionary dynamics, ours, to our knowledge, is the first to introduce the concept of the attractor arc, environment-actuated driven oscillation, trapping zone and escape. We discover an oscillatory fluctuation between labor strategies across market cycles as well as additional transformations resulting from various technology and policy landscapes. 

In this paper, we impart four notable contributions to the cannons of evolutionary dynamics and existing gig literature. First, we introduce a new type of game, replicator dynamics with attractor arcs. We present our model by formalizing our concepts of the attractor arc, environment-actuated driven oscillation, trapping zone and escape. While canonical applications of evolutionary game theory focus on the evolutionary stable strategy (ESS), our model assumes that the system exhibits oscillatory dynamics and cannot fixate on an ESS. In our theoretical extensions, we show how the attractor arc can drift around the phase space and change orientation to reflect evolving labor market composition and dynamic strategy sensitivities. Second, we demonstrate how evolving market conditions effectuate idiosyncratic fluctuations in labor strategies for firms of varying skill sets and size. Here, we find a mismatch in oscillatory behavior for high and low skill firms, which we explain through their differing operational requirements and business sensitivities. Third, we address tensions regarding technology's role in the future of the gig economy. Our findings are consistent with the notion that gig jobs in the early gig economy were elite, high payoff roles such as a senior advisor or management consultant \citep{hyman2018temp}. Our results suggest that technology enabled commodity skill workers to enter and sustainably participate in the gig economy, thereby decreasing average gig payoff and increasing overall gig participation. Regarding future technological advancements, our model presents a theoretical payoff framework informing the possibility of both future growth and stagnation of the gig economy. Fourth, we explore regulatory implications within the gig economy by demonstrating how intervals of lenient and strict policy alter firm and worker sensitivities to different laborer strategies.

\section{Material and methods}

\subsection{Overview.}
In our model, we first explore market influences on firm and laborer strategies in different firm settings. We discretize firms by size and skill set into four classifications : small low-skill firms (\emph{i.e., family owned restaurant business}), large low-skill firms (\emph{i.e., Uber}), small high-skill firms (\emph{i.e., early stage technology startup}) and large high-skill firms (\emph{i.e., Microsoft}). For each of the four firm categories, we generate payoffs that reflect the incentives for each labor strategy in a bear or bull market. We recast gig-employee trade-offs as employment incentives or deterrents for each strategy. For the firm, the strategy set consists of hiring either a gig worker or an employee. For the laborer, the strategy set consists of participating in the labor market as either a gig worker or an employee. To generate each strategy payoff for the firm, we consider factors such as operational revenue, cost of labor and other hiring considerations such as worker reliability, cost of talent acquisition and labor flexibility. For laborer payoffs, we factor in compensation, bonuses and utility gained from alternative engagements outside the work contract. Firm and laborer payoffs are represented in a payoff bi-matrix. We detail our assumptions and payoff generation in Appendix A. We generate 8 final payoff bi-matrices (see Appendix C.1-8) to represent each of our four firm categories in bear and bull markets. 

Once we generate our payoff data, we derive our evolutionary model, replicator dynamics with attractor arcs, and apply the modeling analysis to our generated payoffs. We apply replicator dynamics to model changes in firm and laborer preferences for gig labor across bear and bull markets. Finally, in two theoretical extensions, we detail how changes in payoffs can be applied to study technology and policy leverage in the gig economy.  

\subsection{Evolutionary dynamics of gig economy labor preferences.}
As what follows, we detail the derivation and characteristics of our evolutionary model. First we introduce the replicator equations for 2x2 asymmetric bi-matrix games. By means of two sample bi-matrices, we analyze the phase diagrams and discuss saddle points and initial conditions. Finally, we explore oscillatory dynamics and introduce our theory on the attractor arc, trapping zones, environment-actuated driven oscillation and escape. In following sections, we apply the model to our generated payoffs.

\subsubsection{Replicator Equations for Asymmetric Bi-matrix Games}

In our model, we employ the replicator equation, a differential equation that determines the evolving composition of strategies in a population \citep{hofbauer1998evolutionary, traulsen2009stochastic, zeeman1980population}, to study gig economy labor strategies. In particular, we are interested in how firm and laborer preference for gig labor strategies evolves across market cycless. We provide the general replicator equation where $x_i$ denotes the proportion of strategy type $i$ in the population, $\pi_i$ is the fitness of strategy type i and $\overline{\pi}$ represents the average payoff across the entire population. Fitness of a strategy type can be understood as the expected payoff for that strategy.
\begin{equation}
\dot{x_i} = x_i(\pi_i - \overline{\pi})
\end{equation}

\noindent For asymmetric bi-matrix games, replicator equations take the following form where $\dot{x_i}$ denotes the evolution for player 1 strategies and $\dot{y_i}$ denotes the evolution for player 2 strategies. In our model, player 1 is the laborer and player 2 is the firm. A and B denote the respective payoffs in matrix form for player 1 and player 2. $\Vec{x}$ and $\Vec{y}$ denote the strategies for player 1 and 2 respectively. In vector form, the strategy set for laborers is represented as 
$\Vec{x} = (x_1,x_2)^T$
and the strategy set for firms as
$\Vec{y} = (y_1,y_2)^T$; type 1 strategies typify gig and type 2, employee. Each strategy takes a value in the domain [0,1] and represents the probability the strategy is selected; therefore, $x_1+x_2=1$ and $y_1+y_2=1$.    
\begin{equation}
\dot{x_i} = x_i((A\Vec{y})_i - \Vec{x}\cdot{(A\Vec{y}}))
\end{equation}
\begin{equation}
\dot{y_j} = y_j((B\Vec{x})_j - \Vec{y}\cdot{(B\Vec{x}}))
\end{equation}

\noindent Selection intensity, denoted with $\omega \in [0,1]$, represents the frequency in which firms and laborers interact in the labor market. When firms and laborers do not interact in the labor market, the composition of employees and gig workers remains constant. When firms and laborers choose to participate in the labor market (i.e., firms hiring for and laborers seeking new employment roles), gig and employee decisions are determined based on respective payoff incentives, and the composition of employees and gig workers evolves accordingly. In evolutionary game theory, this social learning process can be modeled as the Moran process \citep{traulsen2009stochastic, de2019fixation}. In our model, $\omega$ constitutes the rate of change for strategy densities in firm and laborer populations. For $\omega$ = 0, the fitness of the strategy type is 0 as the player does not interact in the labor market, and the rate of change for gig-employee strategy densities is 0. When $\omega = 1$, the fitness equates to the payoff for the strategy type, and firms and laborers engage in the labor market at the maximum cadence. We have
\begin{equation}
\pi_i= 1-\omega +{\omega}(A\Vec{y})_i
\end{equation}

\noindent Since each player's strategy set sums to 1, we can mathematically represent our model with just $x_1$ and $y_1$. For 2x2 bimatrix games incorporating selection intensity, replicator equations can be represented in the following form:
\begin{equation}
\dot{x_1} = {\omega}x_1(1-x_1)((A\Vec{y})_1 - (A\Vec{y})_2)\end{equation}
\begin{equation}
\dot{y_1} = {\omega}y_1(1-y_1)((B\Vec{x})_1 - (B\Vec{x})_2)\end{equation}

\noindent Our model involves a pair of GameStates, see Appendix A.1.; GameState pairs consist of a firm category in a bear and bull market. For instance, Small Low Bear and Small Low Bull make up a GameState pair that portrays a small low skill firm in bear and bull markets. Subscripts $l$ and $f$ denote laborer and firm payoffs respectively. We append 0 and 1 to the payoff subscripts to denote bear and bull market GameStates respectively. 
\begin{figure*}[h!]
    \centering
    \makebox[\textwidth]{
    \centering
    \includegraphics[width=14cm]{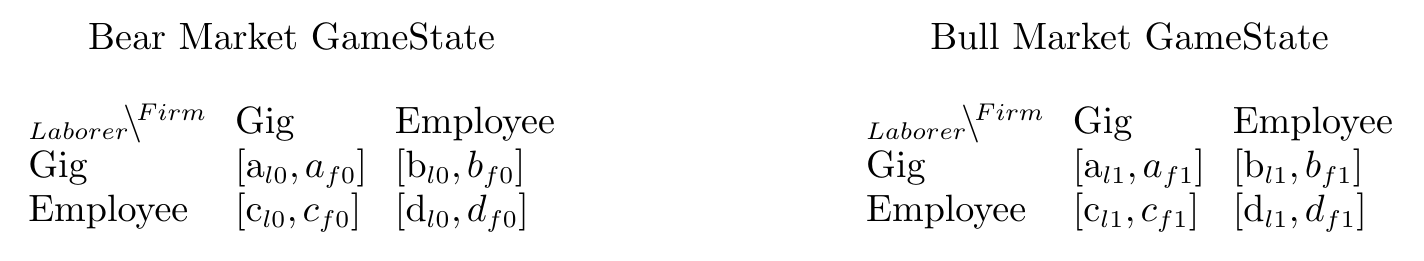}
}
\end{figure*}

We reconstitute our pair of GameState matrices, disjoining firm and laborer payoffs.
\begin{equation}
L_{Bear} = \begin{pmatrix}
a_{l0} & b_{l0} \\
c_{l0} & d_{l0}
\end{pmatrix}
\quad
L_{Bull} = \begin{pmatrix}
a_{l1} & b_{l1} \\
c_{l1} & d_{l1} \\
\end{pmatrix}
\end{equation}

\begin{equation}
F_{Bear} = \begin{pmatrix}
a_{f0} & b_{f0} \\
c_{f0} & d_{f0} 
\end{pmatrix}^T
\quad
F_{Bull} = \begin{pmatrix}
a_{f1} & b_{f1} \\
c_{f1} & d_{f1}
\end{pmatrix}^T
\end{equation}

An environment coefficient, $n \in [0, 1]$, represents market condition. $n = 0$ denotes the bear market and $n = 1$ denotes the bull market. $n$ can take any value
between 0 and 1; for instance, $n = 0.5$ signifies that the environment is a neutral
market, the midway point in a transition between bear and bull market conditions. Applying this environment coefficient, we rephrase our firm and laborer
payoffs to account for the domain of market conditions.
\\ 

\noindent\begin{equation*}
A(n)=L_{General} = \begin{pmatrix}
(1-n)a_{l0}+na_{l1} & (1-n)b_{l0}+nb_{l1}\\
(1-n)c_{l0}+nc_{l1} & (1-n)d_{l0}+nd_{l1}\\\end{pmatrix}
\end{equation*}
\vspace{3mm}
\noindent\begin{equation*}
B(n)=F_{General} = \begin{pmatrix}
(1-n)a_{f0}+na_{f1} & (1-n)b_{f0}+nb_{f1}\\
(1-n)c_{f0}+nc_{f1} & (1-n)d_{f0}+nd_{f1}\\\end{pmatrix}^T
\end{equation*}

\vspace{3mm}
\noindent We apply our general firm and laborer payoffs to our replicator equations and conclude our derivation.
\begin{equation}
\begin{split}
\dot{x_1} = {\omega}x_1(1-x_1)&(([(1-n)a_{l0}+na_{l1}]y_1+[(1-n)b_{l0}+nb_{l1}](1-y_1)) \\ & -([(1-n)c_{l0}+nc_{l1}]y_1+[(1-n)d_{l0}+nd_{l1}](1-y_1)))  
\end{split}
\end{equation}

\begin{equation}
\begin{split}
\dot{y_1} = {\omega}y_1(1-y_1)&(([(1-n)a_{f0}+na_{f1}]x_1+[(1-n)c_{f0}+nc_{f1}](1-x_1)) \\ & -([(1-n)b_{f0}+nb_{f1}]x_1+[(1-n)d_{f0}+nd_{f1}](1-x_1)))
\end{split}
\end{equation}

\section{Results}

\subsection{Key Concepts and Theoretical Analysis of the Evolutionary Game Theory Model}

\subsubsection{System Equilibria}
In Appendix B.1, we solve for our evolutionary system's fixed points for the general case. For each fixed point, we examine the stability of the equilibrium by analyzing the eigenvalues of the Jacobian matrix. We find that our system has two stable fixed points at $(0,0)^*$ and $(1,1)^*$, two unstable fixed points at $(0,1)^*$ and $(1,0)^*$, and a saddle point whose position depends on firm and laborer payoff values. 

\subsubsection{Saddle Points}

We provide analysis for the saddle point with a theoretical GameState pair. For simplification purposes, we assign all mismatching strategies a payoff of $0$ as mismatching strategies take marginal values in respect to matching strategies (we refer to Appendix B for the theoretical analysis of general cases). 
\begin{figure}[h!]
    \centering
    \makebox[\textwidth]{
    \centering
     \includegraphics[width=14cm]{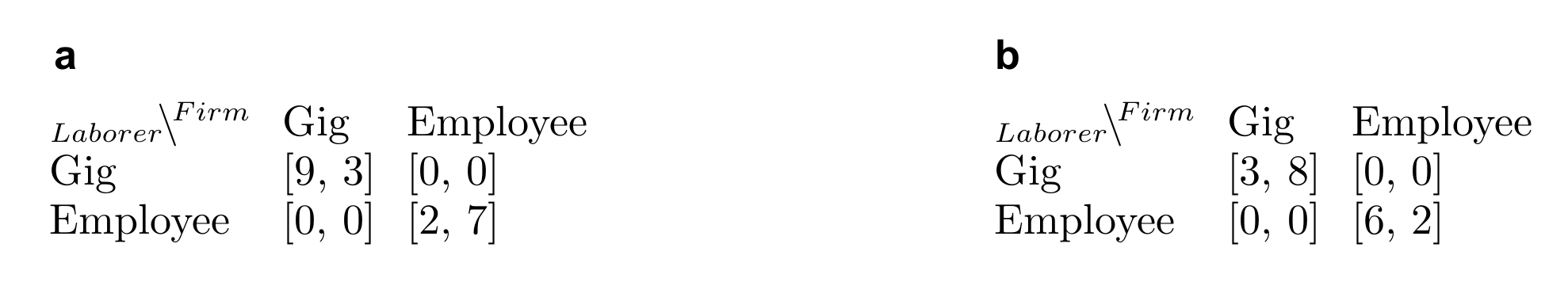}
     }
     \caption{Theoretical GameState Pair Payoff Matrices Used in Demonstrations. (a) Bear Market GameState (b) Bull Market GameState}
\end{figure}

\noindent In the bear market GameState, $a_l > d_l$ and $a_f < d_f$. The laborer receives a higher payoff for competing as a gig worker (payoff: 9 vs. 2) and the firm receives a higher payoff for hiring an employee (payoff: 3 vs. 7). 

In the bull market GameState, $a_l < d_l$ and $a_f > d_f$. The laborer receives a higher payoff for competing as an employee (payoff: 3 vs. 6) and the firm receives a higher payoff for hiring a gig worker (payoff: 8 vs. 2).

\subsubsection{Saddle Point Geographies}

In our phase diagrams, $y_1$ denotes firm strategy for gig and $x_1$ denotes laborer strategy for gig, consistent with our replicator equations.

\begin{figure}[h!]
    \centering
    \makebox[\textwidth]{
    \includegraphics[width=17cm]{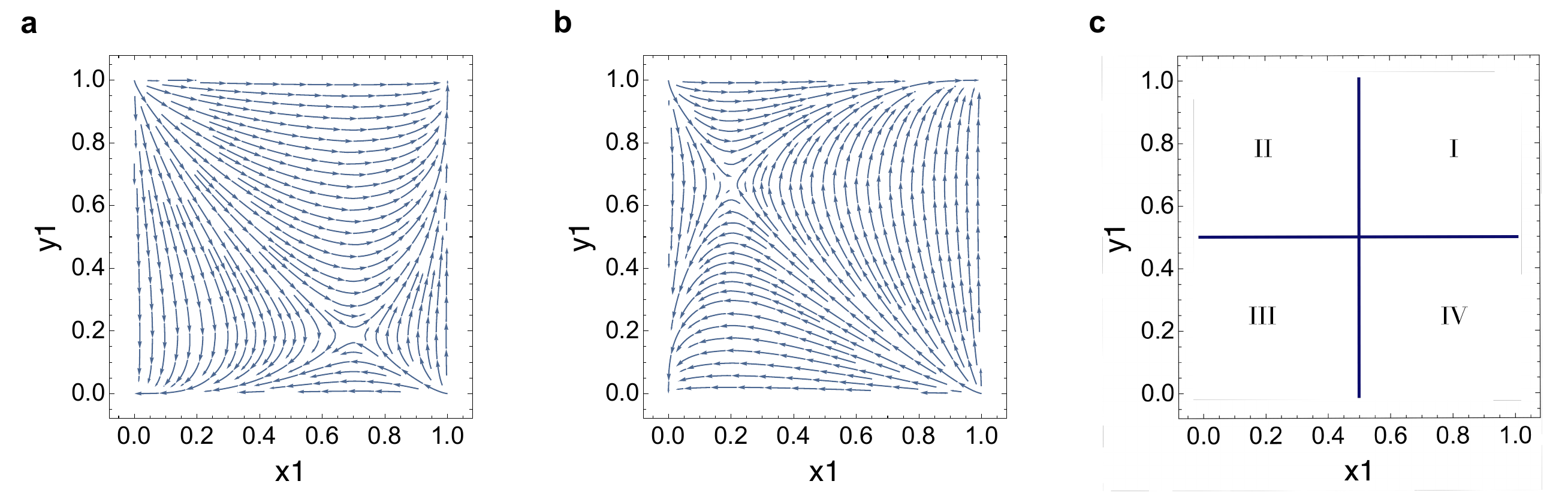}
    }
    \caption{Saddle Point Geographies with Theoretical GameState Payoffs, See Figure 1. (a) Bear GameState $n=0$ (b) Bull GameState $n=1$ (c) Quadrant Legend}
\end{figure}

\noindent Payoff relationships determine the geography of the saddle point. We list the general conditions for saddle point positions in regard to our quadrant legend.  
\\

\noindent Quadrant I: $a_l < d_l$ and $a_f < d_f$\\
Quadrant II: $a_l < d_l$ and $a_f > d_f$\\
Quadrant III: $a_l > d_l$ and $a_f > d_f$\\
Quadrant IV: $a_l > d_l$ and $a_f < d_f$\\

 \noindent Indeed, the saddle point for the bear GameState is located at ($\frac{7}{10}$,$\frac{2}{11}$) in quadrant IV. The saddle point for the bull GameState sits at ($\frac{1}{5}$,$\frac{2}{3}$) in quadrant II. 

\subsubsection{Attractor Arc, Driven Oscillation and Trapping Zones}

\paragraph{Attractor Arc.} In our model, we refer to our model's saddle point as an \emph{attractor} (more strictly, which acts as an attractor for some trajectories and a repellor for others), a term we adopt and extend from the mathematical study of dynamical systems which describes a locale in the phase space that the system gravitates towards \citep{milnor1985concept, eckmann1985ergodic}. 

For a dynamical system with an environment $n$ that does not change states as a function of time, $\dot{n}=0$, the system will evolve to one of the two stable equilibria at $(0,0)^*$ or $(1,1)^*$ dependent on initial condition; the system represents the composition of gig strategies in firm ($y$) and laborer ($x$) populations. We demonstrate this concept with $n=0$, denoting the bear market GameState, and initial conditions ($\frac{1}{4}$,$\frac{1}{4}$) and ($\frac{3}{4}$,$\frac{3}{4}$) to show two evolutionary paths. 
\begin{figure*}[h!]
    \centering
    \makebox[\textwidth]{
    \includegraphics[width=5cm]{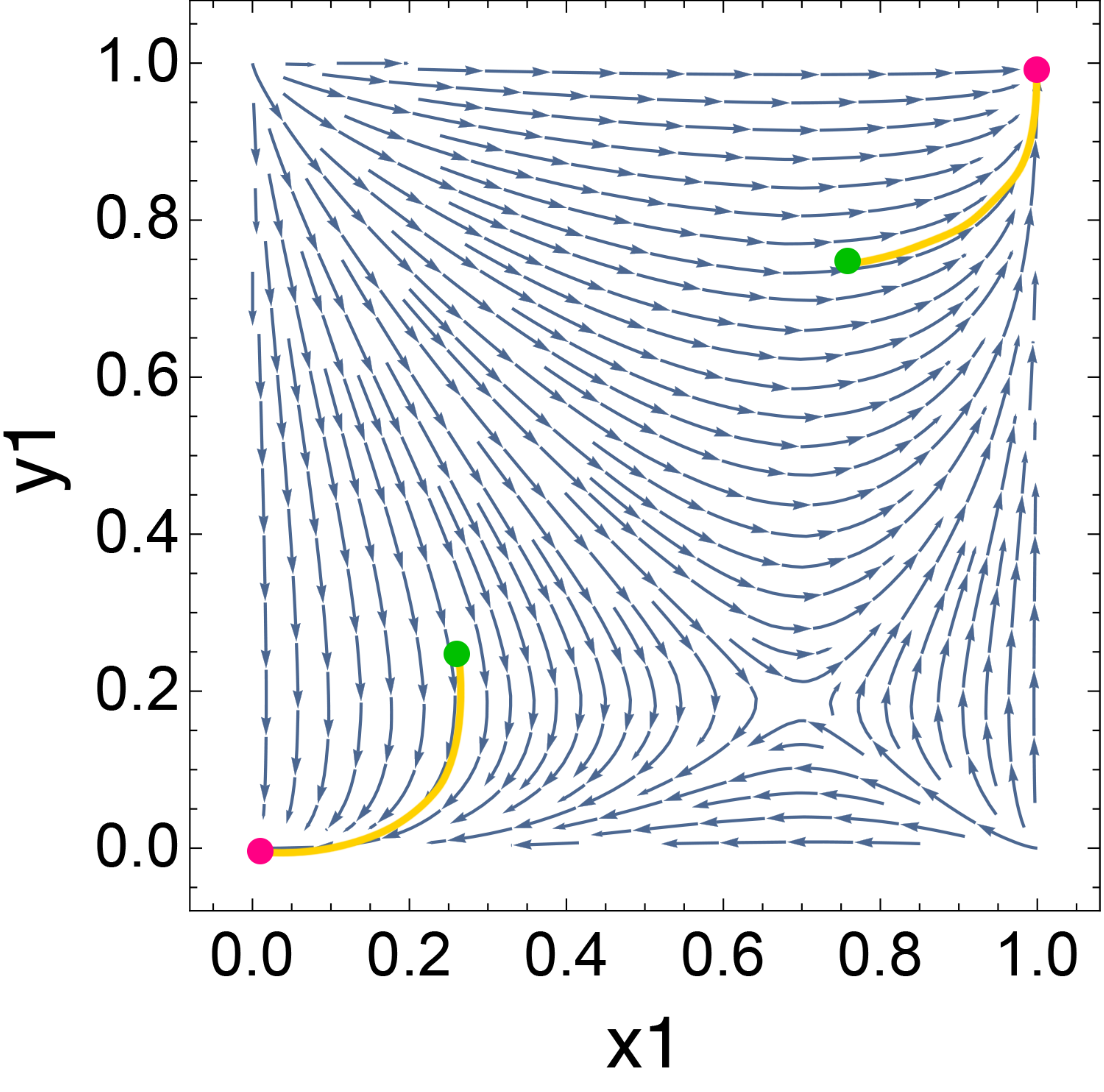}
    }
         \caption {Evolutionary Behavior for $n=0$, $\dot{n}=0$ with Theoretical GameState Payoff, see Figure 1. In this visualization, green represents initial condition, yellow represents the evolutionary path and red represents the final system position at an ESS. Two evolutionary journeys are visualized with initial conditions ($\frac{1}{4}$,$\frac{1}{4}$) and ($\frac{3}{4}$,$\frac{3}{4}$).}
 \end{figure*}

For a dynamical system with an environment $n$ that evolves as a function of time, $\dot{n}\neq 0$, phenomena of interest is centered around the \emph{attractor arc}. The attractor arc represents the entirety of possible attractor (saddle point) positions given $n\in[0,1]$. Mathematically speaking, the attractor arc defined in the present work can be viewed as an invariant manifold; if starting from one given point at the attractor arc, the system's trajectories under changing market conditions will remain on the arc. To graphically represent the attractor arc, we superimpose our theoretical bear, $n=0$, and bull, $n=1$, GameState phase diagrams and plot the saddle points for all $n\in[0,1]$. The phase diagram for $n=0$ is superimposed in orange while that of $n=1$ is superimposed in blue. Below, the attractor arc is represented in purple. It is important to note that while this superimposed visual exhibits five reference saddle points, only one saddle point exists at any given time $t$.

\begin{figure*}[h!]
    \centering
    \makebox[\textwidth]{
    \includegraphics[width=5cm]{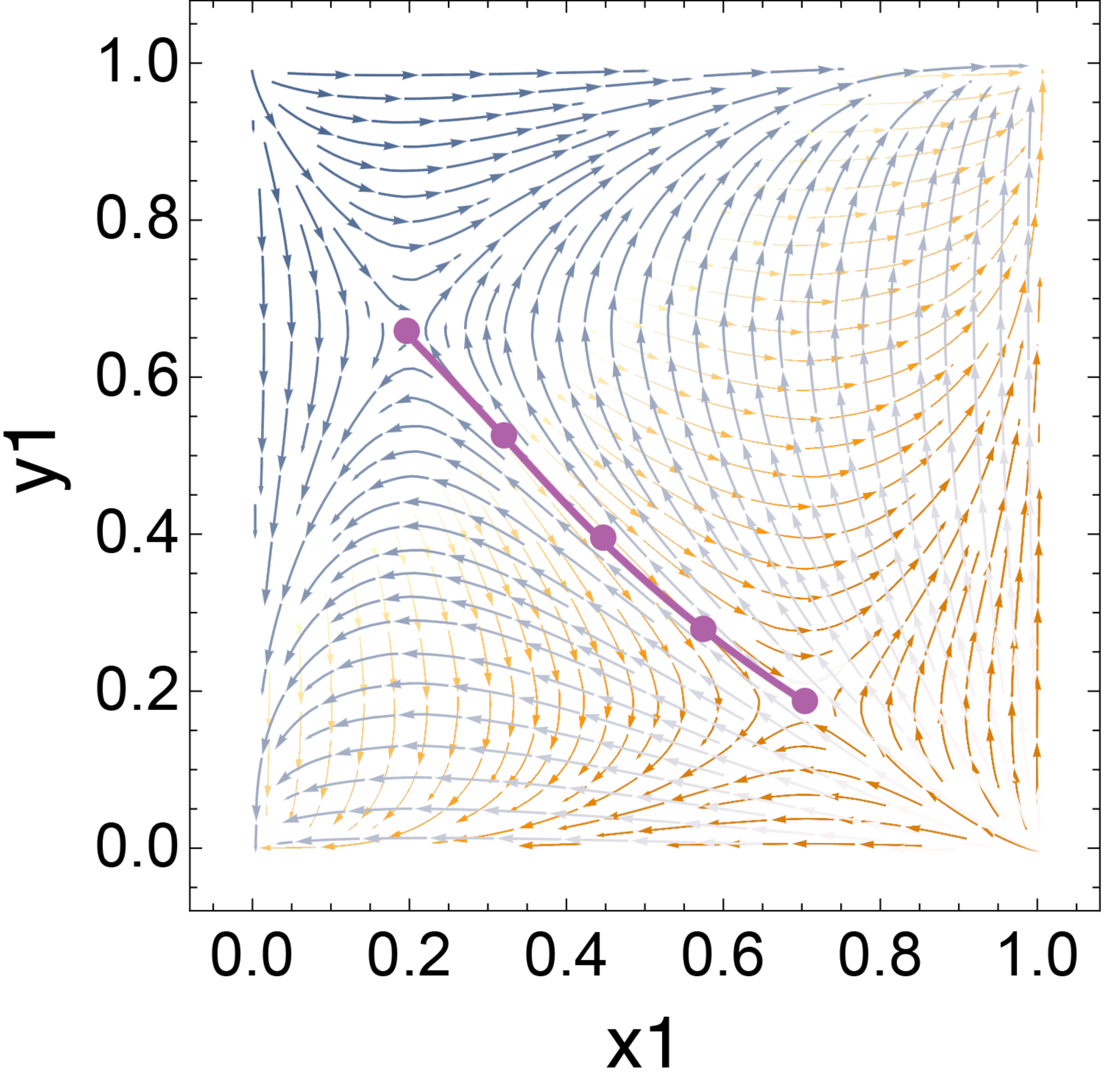}
    }
        \caption {2D Attractor Arc Mapping on Superimposed Theoretical GameState Payoff when $n=0$ and $n=1$, see Figure 1. The attractor arc represents the entirety of possible attractor positions given $n\in[0,1]$. Reference points on the attractor arc demonstrate attractor positions when $n=0$, $n=0.25$, $n=0.5$, $n=0.75$ and $n=1$. }

 \end{figure*}
We attain the preceding arc (see Figure 4) by collapsing the three dimensional [$x_1$,$y_1$,$n$] attractor arc (see Figure 5) onto $x_1$ and $y_1$ dimensions. 
\begin{figure*}[h!]
    \centering
    \makebox[\textwidth]{
    \includegraphics[width=10cm]{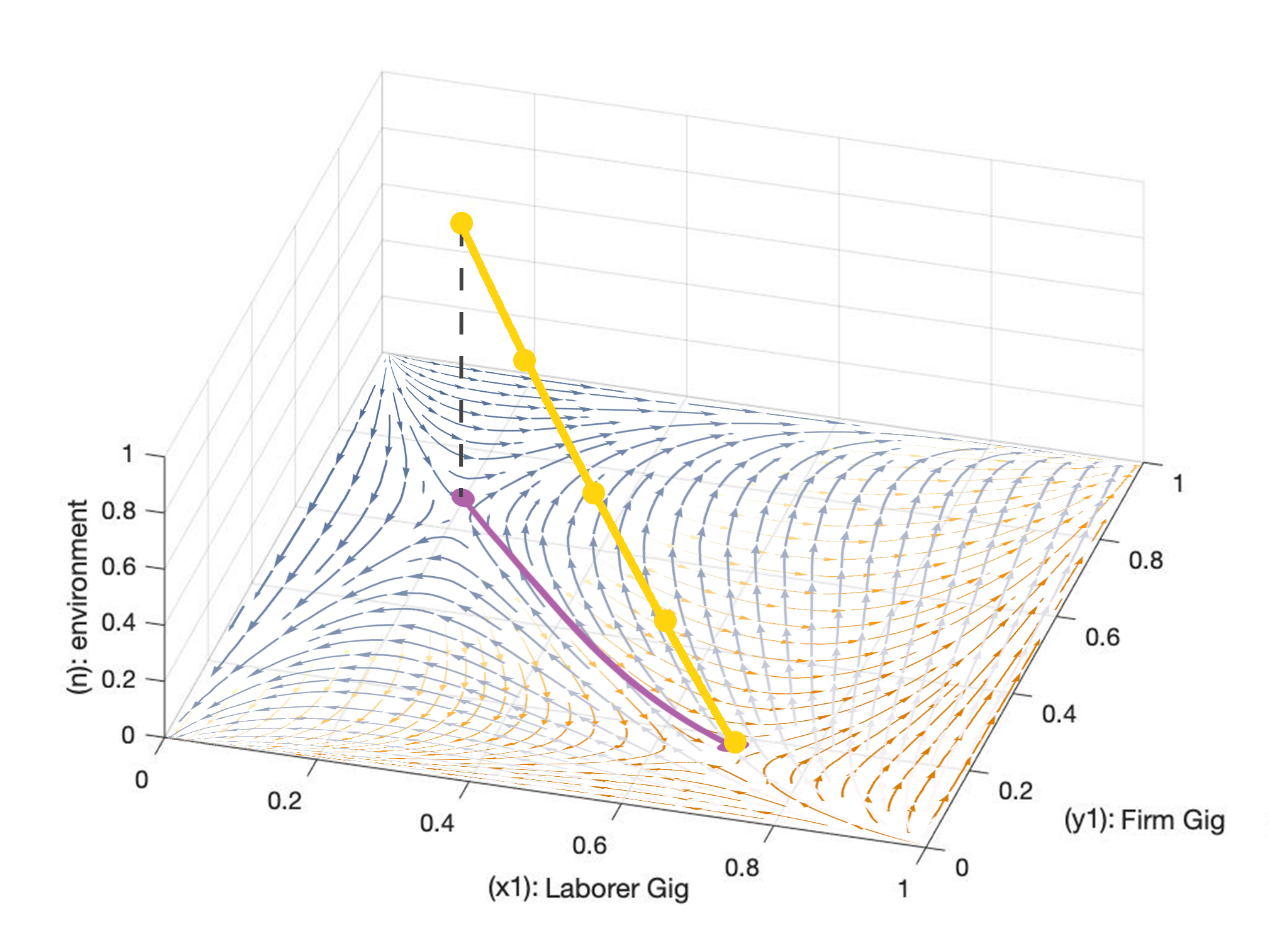}
    }
         \caption {3D Attractor Arc. The 3D arc is represented in yellow with reference attractor positions when $n=0$, $n=0.25$, $n=0.5$, $n=0.75$ and $n=1$. The projected 2D arc is represented in purple, consistent with the antecedent diagram, see Figure 4.}
 \end{figure*}
 
For our demonstrations, we apply a simple step-wise function for $\dot{n}$ such that the environment instantaneously alternates between $n=0$ and $n=1$ every 5 time units, see Figure 6. For clarity, we plot our selected $\dot{n}$ to help visualize the rate of change for the environment. Notably, our step-wise $\dot{n}$ implies that the attractor will jump from the two extremes of the attractor arc corresponding to $n=0$ and $n=1$. While we provide a reference attractor arc in all demonstrations, our $\dot{n}$ implies the attractor will not take an intermediary position on the arc.
\begin{figure}[h!]
    \makebox[\textwidth]{

    \centering
    \includegraphics[width=15cm]{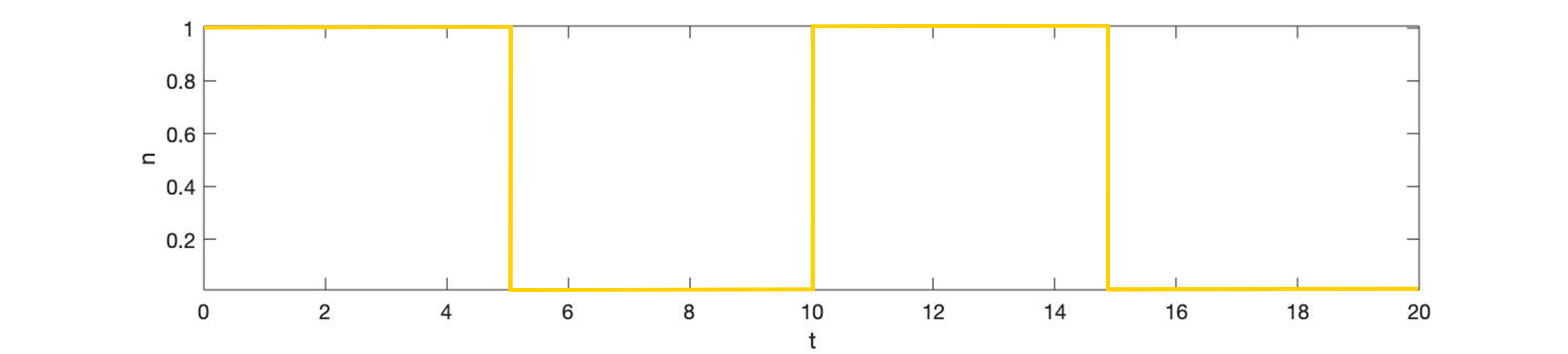}
    }
 \caption {Chosen $\dot{n}$, n as a Function of t, to be used in Demonstrations}
 \end{figure}
\subsubsection{Shepherding Attractors, Driven Oscillation and Trapping Zones}

For a given pair of GameStates, $\dot{n}$ determines the orbit and moving speed of the attractor. As the attractor orbits the attractor arc, the attractor's oscillation can drive the system to oscillate as well. We refer to this as a driven oscillation. Near the attractor arc, there exists a trapping zone where the system can oscillate for numerous periods. Here, the attractor has a shepherding role. In order for the attractor to herd the system for numerous periods, $\omega\neq0$ must be small enough compared to $\dot{n}\neq0$ such that the system does not escape the ends of the attractor arc. A simple analogy can help elucidate this concept. The attractor behaves as a shepherd who can only move along one line, the attractor arc. The system behaves like a sheep that is running towards or away from the shepherd, depending on the orientation of the attractor arc. The shepherding attractor must move from one end of the arc to the other faster than the sheep in order to trap it. If the sheep reaches an escape boundary such that the shepherding attractor can not keep up, it will escape and end up at one of the two stable equilibria at $(0,0)^*$ or $(1,1)^*$. Escape from the trapping zone depends on the nontrivial relationship between $\dot{n}$ and $\omega$. Therefore, given $\omega$ is very small such that the system is evolving much slower than the attractor, the trapping zone behaves as a pseudo-stable equilibrium between a pair of GameStates. Without environmental changes, the system remains stationary at the attractor arc, and it will slowly escape to one of the stable equilibria.  

\begin{figure*}[h!]
    \centering
    \makebox[\textwidth]{
    \includegraphics[width=17cm]{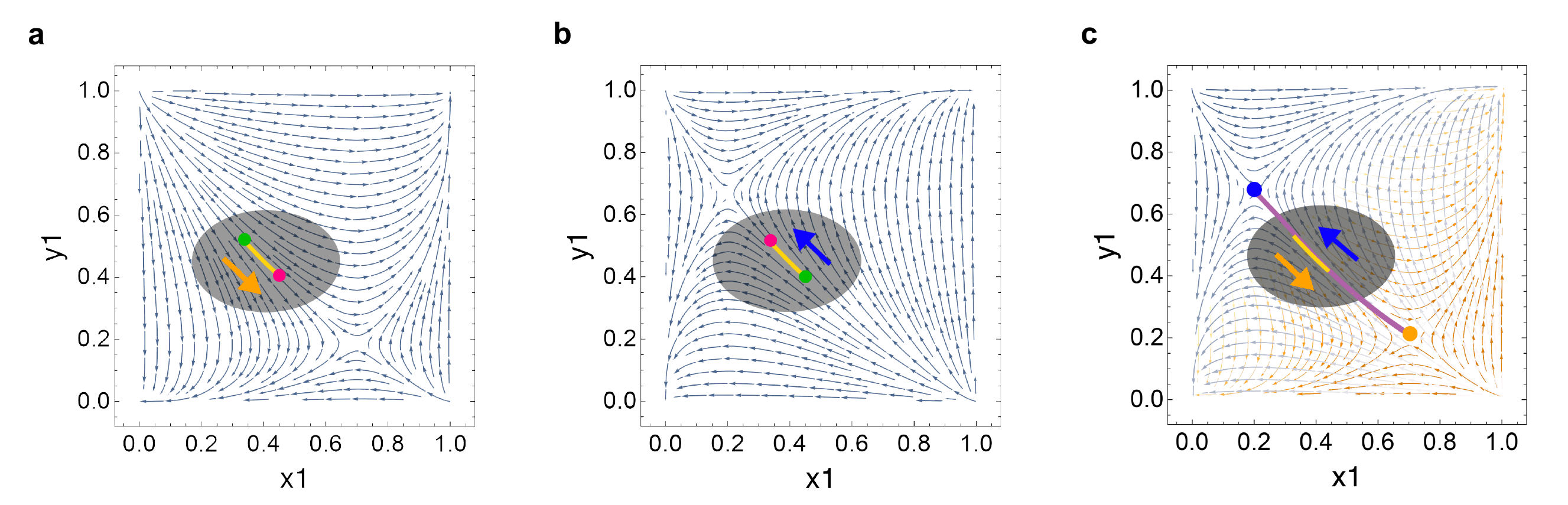}
    }
    \caption{Concept Visuals: Shepherding Attractors and Driven Oscillation. (a) Evolution in Bear Market (b) Evolution in Bull Market (c) Driven Oscillation. In (a) and (b), we plot the evolutionary trajectories for a bear and bull market. For each phase diagram, green denotes initial condition, red denotes ending destination, and yellow denotes the evolutionary path. In (c), a reference attractor arc is plotted in purple and attractor positions at $n=0$ and $n=1$ are represented in orange and blue respectively. The trapping zone orbit is plotted in yellow. The opaque black ellipse is a background element to help visually contrast with the trapping zone. This oscillation models $\omega=0.5$ and initial conditions $n=1$ and $(0.45, 0.4)$, attractor position when $n=0.5$. In this figure, we use a relatively large $\omega$ for the purpose of visualizing the evolution in (a) and (b). }
\end{figure*}

\paragraph{Escape and Implications}
Assuming that the system has previously existed by oscillating in tescape boundaryhe trapping zone, escape is possible if there is a perturbation that changes $\dot{n}$ and or $\omega$ such that the system reaches escape boundary. Once the system reaches escape boundary, the system will eventually escape the trapping zone to one of the stable equilibria at $(0,0)^*$ or $(1,1)^*$, see Appendix B.2.3.

With escape, it is important to note that initial condition is crucial in determining which stable equilibrium the system escapes to. If $\omega$ increases twenty-fold such that the system reaches escape boundary at the start of a bear market, $n=0$, rather than at the start of a bull market, $n=1$, the system evolves to $(0,0)^*$ rather than $(1,1)^*$. A claim based on which of the two stable equilibria the system escapes to is indefensible, as this result is subject to the initial conditions. As such, we theorize the possibility of escape but do not run our models to make a claim for a specific escape destination. Therefore, we can only conclude that changes in $\dot{n}$ and $\omega$ can allow the system to reach escape boundary and result in an accelerated escape to one of the two stable equilibria. However, we can not conjecture which stable equilibrium the system escapes to.

\paragraph{Selection of Initial Conditions.} When applying this model, it is unfitting to prepare any arbitrary initial condition because different initial conditions can result in different evolutionary outcomes, see Appendix B.2.3. Therefore, all findings or claims fixating on a specific ESS can be countered with the selection or preparation of another initial condition. 

In our model, we presume that some co-existence of gig and employee strategies has always been present in the labor market. Our evolutionary system informs us that if the labor market consisted of only one type of worker (gig or employee) in the past, there would be no co-existence of gig and employee strategies today, as the system would have remained fixated on that ESS; therefore, we reason that the present day co-existence of gig and employee strategies necessitates a historical co-existence of gig and employee strategies. 

Mathematically, this implies that our system has always been ``trapped'' in a state of oscillatory dynamics up until the observable present-day. Appropriately, in this work, we have defined the mechanism that ``traps'' the system in this pseudo-stable state of gig and employee co-existence. 

It is sensible for our system to evolve within the pseudo-stable trapping zone as this represents the present-day domain of oscillatory dynamics and observable co-existence of gig and employee strategies. Therefore, any point in the trapping zone is a suitable initial condition. In our models, we use the attractor position at $n=0.5$, the midway point between a bear and bull market transition, as an estimator for a point in the trapping zone. 

\paragraph{Attractor Arc Drift and Tilt.} If we consider payoffs as a function of time, $\dot{A}, \dot{B} \neq0$, the attractor arc itself will evolve, see Figure 8. Accordingly, this implies the trapping zone will change position with the attractor arc because the system's orbit is a driven oscillation. Assuming the system exists by always oscillating in the pseudo-stable trapping zone, evolving payoffs can help explain how the \emph{system's orbit}, an orbit in the trapping zone, can move around the phase space. The shape and orientation of the arc at any given time $t$ depends on $\dot{A}$ and $\dot{B}$. In later sections, we investigate payoff operations that cause the attractor arc to drift (change position in the phase space) and tilt (change orientation in the phase space). 

\begin{figure*}[h!]
    \centering
    \makebox[\textwidth]{
    \centering
\includegraphics[width=12cm]{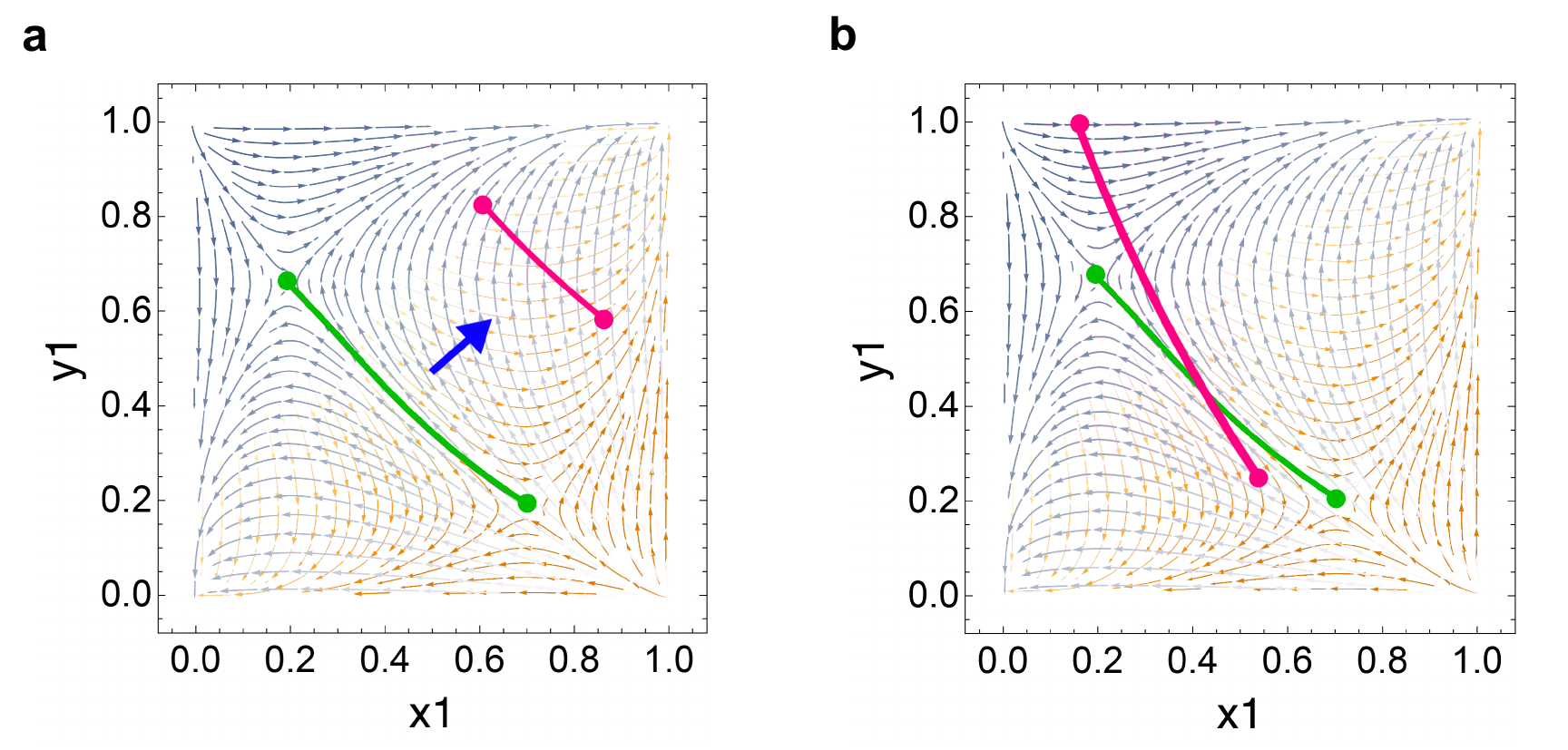}
    }
     \caption{Concept Visuals: Attractor Arc Transformation. (a) Attractor Arc Drift $\dot{A}, \dot{B}\neq0$ (b) Attractor Arc Tilt $\dot{A}, \dot{B}\neq0$. In (a), the green arc applies the Theoretical GameState Pair Payoff, see Figure 1, and the red arc applies a High Employee Payoff Matrix Operation, see Figure 13. In (b), the green arc applies the Theoretical GameState Pair Payoff, see Figure 1, and the red arc applies a Lenient Policy Matrix Operation, see Figure 16. }
\end{figure*}

\paragraph{Applications of model to generated payoffs.} Regarding the gig economy, we assume that observable fluctuations in labor strategies reflect the system oscillating within the trapping zone (i.e., what we observe is pseudo stable state at all times). In the following, we investigate the attractor arc and trapping zone patterns for each of our data generated firms (Small Low Skill, Large Low Skill, Small High Skill, Large High Skill), applying our model to the payoff matrix pairs in Appendix C. We hypothesize that the system will never escape the trapping zone, implying that there will always be some co-existence of gig workers and employees. 

\subsection{Market Influences on Firm and Laborer Gig Preference}
%
%
%
%
%
%
%
%
%
%
%
%
%

\subsubsection{Assumptions and Observations.}
In our data generated model, we assume that the system exists by oscillating in the trapping zone. The system's oscillatory behavior reflects observable fluctuations in gig economy strategies across market cycles. For small firms, we assign an $\omega=0.00000001$ that is 50x larger than that for large firms, $\omega=0.0000000002$; since small firms employ a smaller work force, labor composition is more notably impacted by each individual labor decision, implying a faster rate of system evolution. Across our four firm categories, we observe two patterns in oscillatory behavior marked by firm skill set. While low-skill firms and laborers demonstrate a matching oscillatory behavior (when firm preference for gig work increases, laborer preference for gig strategies also increases), high skill firms and laborers exhibit a mismatching oscillatory behavior. Further, the position of the attractor arc is higher on the $y_1$ axis for low-skill firms, implying that these firms maintain a higher proportion of gig workers. In this section, we explore why skill set bifurcates our firm cohort into two categories of oscillatory behaviors. 

Our theoretical firm categories can be mapped to empirical examples. For instance, a small family owned restaurant business can be understood as a small low-skill firm. A large low-skill business embodies ride share companies like Uber and Lyft. An early stage technology startup characterizes a small high-skill firm. Enterprises like Microsoft or Google can be presumed to be large high-skill firms.

\begin{figure*}[h!]
\makebox[\textwidth]{
\centering

\includegraphics[width=13cm]{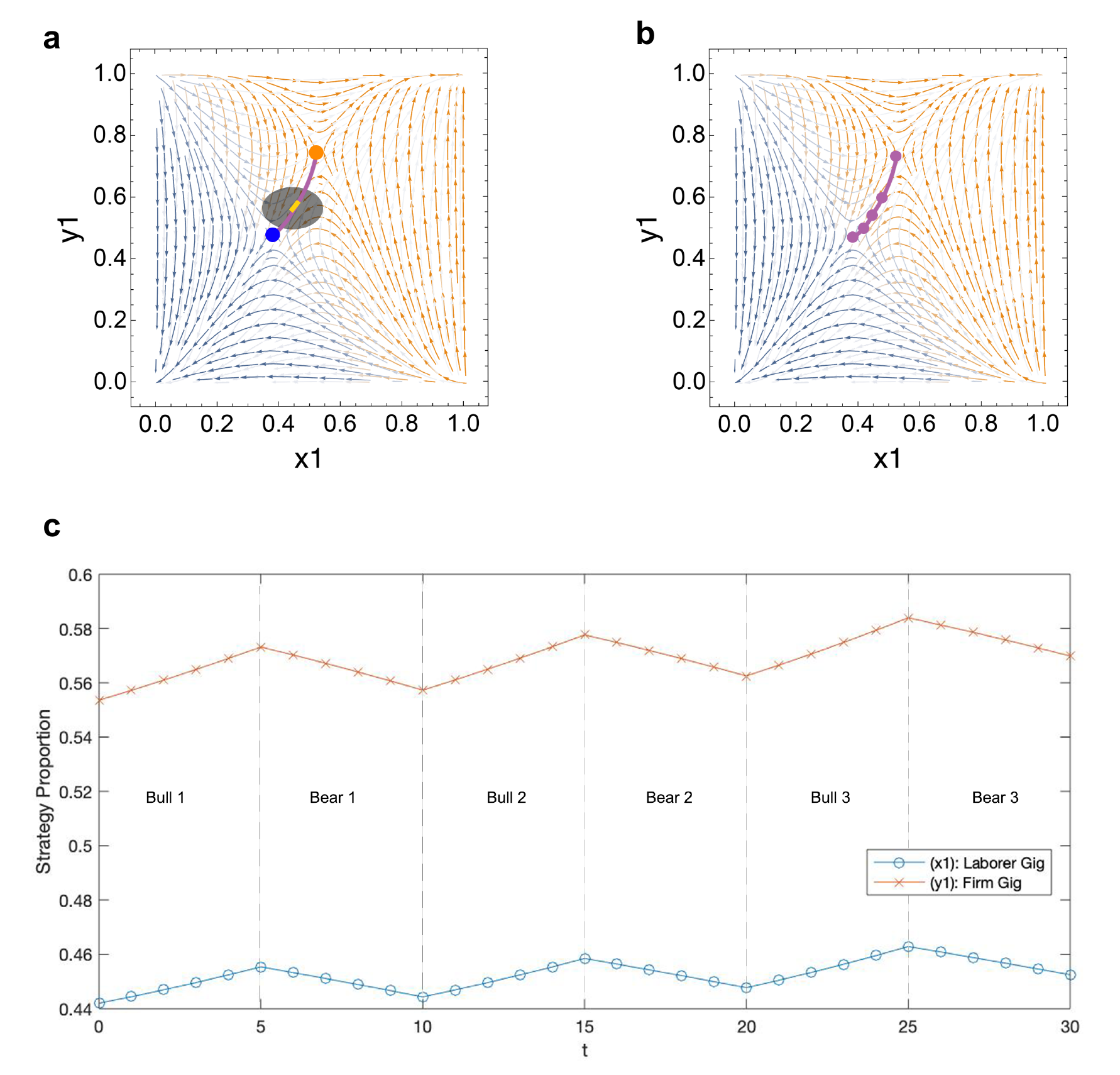}
}
 \vspace*{2mm}\caption{Evolution of Strategy Densities for Small Low-Skill Firm with Initial Conditions $(0.4417, 0.5554)$, the attractor position at $n=0.5$, an approximation for a point in the trapping zone; $\omega = 0.00000001$; and Payoff Matrices Small Low Bear and Small Low Bull, see Appendix C.1 and C.2. (a) Trapping Zone Orbit (b) Attractor Arc (c) Labor Strategy Oscillation Over Three Market Periods. In (a), the trapping zone orbit is plotted in yellow, and attractor positions at $n=0$ and $n=1$ are represented in orange and blue respectively. In (b), we plot a reference attractor arc in purple with attractor positions when $n=0$, $n=0.25$, $n=0.5$, $n=0.75$ and $n=1$. (c) visualizes the fluctuation in firm and laborer preferences for gig strategies over three market periods.}
\end{figure*}

\subsubsection{Firm-Level Discussion}

\paragraph{Market Influence on Low-skill Firms and Laborers}
During a bear market, low-skill firms and their laborers increase their preference for the employee strategy, see Figure 9,10. Several studies find that low-skill laborers are the most adversely affected during a bear market, and they often constitute the majority of layoffs \citep{levine2009labor, uchitelle2009despite, kaye2010impact}. Workers with commodity skills have the fewest options for alternative engagements during this recessionary period. Indeed, the emergence of economic contraction entails a decreased demand for consumer goods and services \citep{levine2009labor}. Present-day gig workers recognize that the structural forces of economic recessions restrict their autonomy in flexible scheduling as service demand abates \citep{lehdonvirta2018flexibility}. In addition to experiencing decreased autonomy, it is reasonable that low-skill laborers increase their preference for employee roles which come with additional financial stability and labor protections.  As low-skill gig labor generally supports operations regarding consumer goods and services, decreased demand for such services may implicate decreased demand for low-skill gig labor to fulfill service operations. Therefore, it is sensible that low-skill firms decrease their demand for gig labor during a bear market. 

\begin{figure*}[h!]
\makebox[\textwidth]{
\centering
\includegraphics[width=13cm]{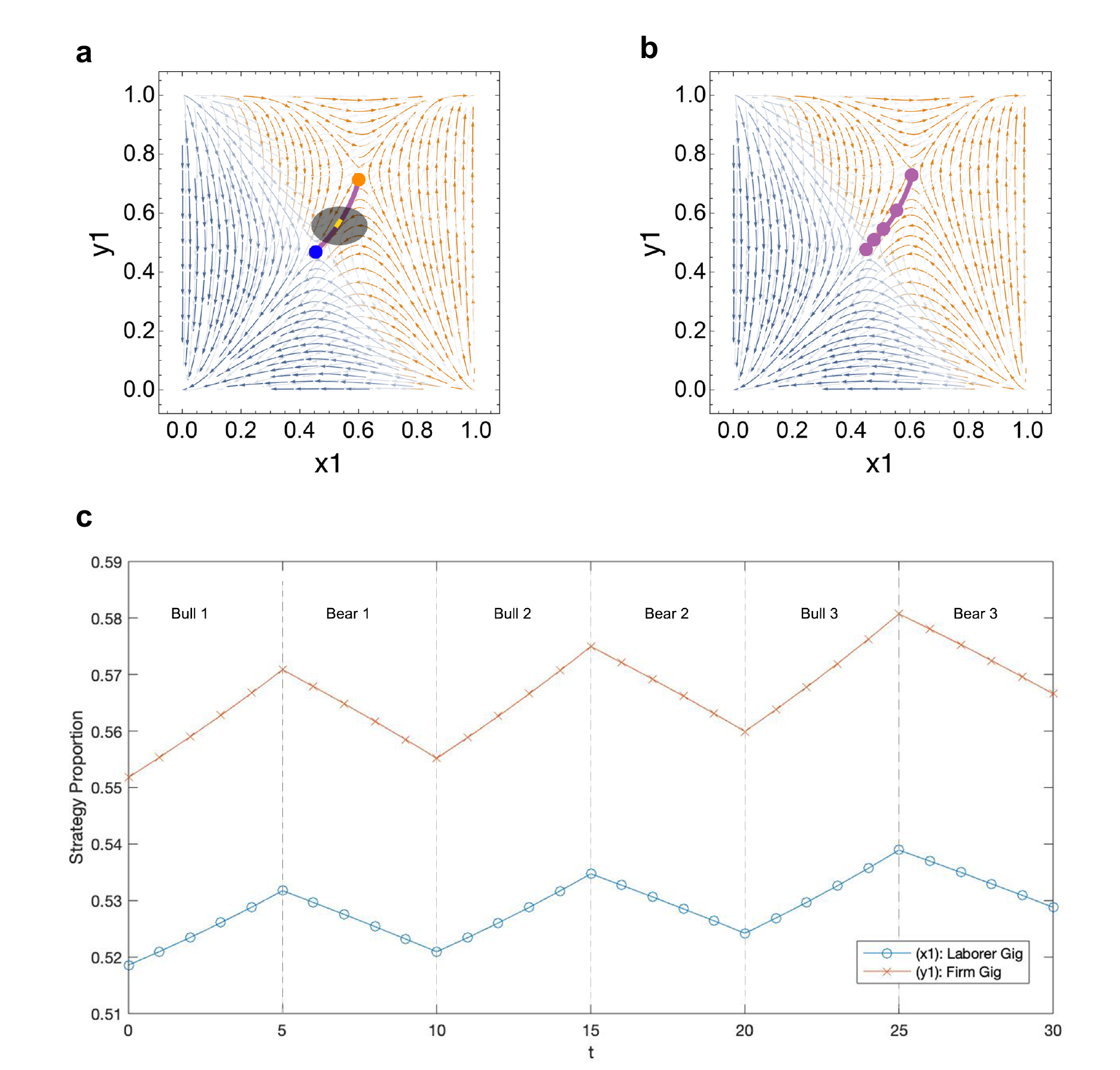}

}
 \vspace*{2mm}\caption{Evolution of Strategy Densities for Large Low-Skill Firm with Initial Conditions $(0.5186, 0.5535)$, the attractor position at $n=0.5$, an approximation for a point in the trapping zone; $\omega=0.0000000002$; and Payoff Matrices Large Low Bear and Large Low Bull, see Appendix C.3 and C.4. (a) Trapping Zone Orbit (b) Attractor Arc (c) Labor Strategy Oscillation Over Three Market Periods. In (a), the trapping zone orbit is plotted in yellow, and attractor positions at $n=0$ and $n=1$ are represented in orange and blue respectively. In (b), we plot a reference attractor arc in purple with attractor positions when $n=0$, $n=0.25$, $n=0.5$, $n=0.75$ and $n=1$. (c) visualizes the fluctuation in firm and laborer preferences for gig strategies over three market periods.}
\end{figure*}

When the market environment changes to a bull market, low-skill firms and laborers increase their preference for the gig strategy, see Figure 9,10. As market optimism rises, demand for consumer goods and services grows. For laborers pursuing gig roles, this implies additional financial stability as demand for services stabilizes \citep{lehdonvirta2018flexibility, broughton2018experiences}. Uber drivers, for instance, can complete more rides each day as a result of increased rider demand. Since the salary of a low-skill gig worker is directly impacted by the number of tasks completed, more rides implies higher compensation. Accordingly, it is logical that low-skill laborers increase their preference for gig strategies during a bull market to capture this increased demand for consumer goods and services. Correspondingly, low-skill firms must adjust their labor demands to accommodate this interval of increased consumerism. It is therefore reasonable that low-skill firms increase their demand for gig labor as they accelerate service operations during bull market conditions. 

\paragraph{Market Influence on High-skill Firms and Laborers} 

In bear market conditions, high-skill firms increase preference for employees while high-skill laborers increase preference for gig work, see Figure 11,12. While recessionary economic conditions are unsympathetic to low-skill laborers, laborers with specific skill sets are more impervious to the impacts of an economic downturn \citep{uchitelle2009despite, kaye2010impact}. Therefore, high-skill laborers have increased leverage during bear market conditions. Coupled with increased flexibility and the opportunity to take on low-skill contracts \citep{fogg2011rising}, it is reasonable that high-skill laborers increase their preference to work gig contracts during bear markets. Conversely, high-skill firms are particularly sensitive to talent retention costs and worker reliability. It is plausible, for example, that firms can expect lower employee churn during a bear market as workers who choose to participate as employees want to retain their positions. Additionally, the cost of error can be particularly detrimental to a high-skill firm during a bear market; therefore, it is sensible that the reliability of an employee over that of a gig worker is more valuable to a high-skill enterprise during a bear market.

\begin{figure*}[h!]
\makebox[\textwidth]{
\centering
\includegraphics[width=13cm]{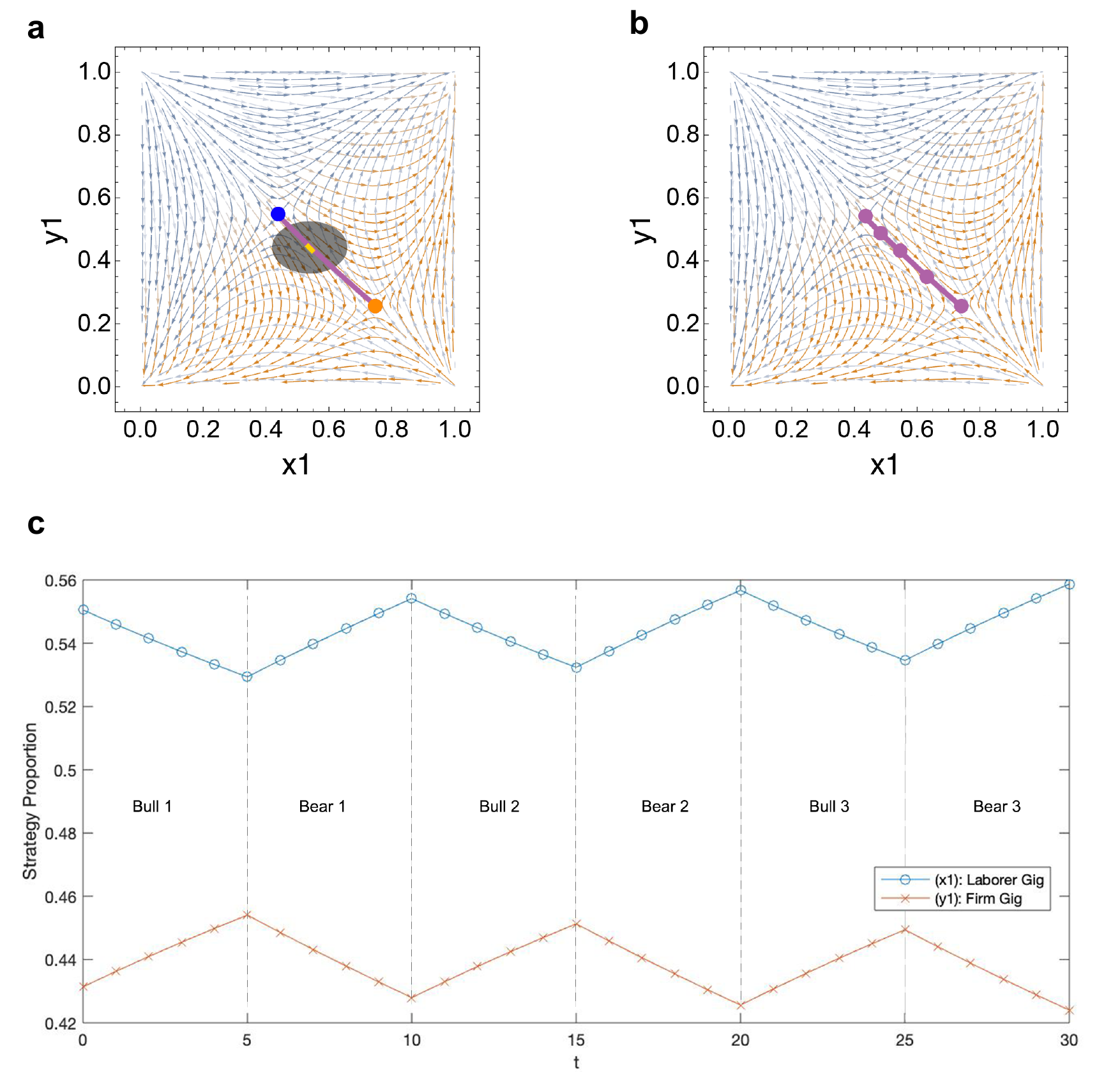}

}
 \vspace*{2mm}\caption{Evolution of Strategy Densities for Small High-Skill Firm with Initial Conditions $(0.5498, 0.4298)$, the attractor position at $n=0.5$, an approximation for a point in the trapping zone; $\omega=0.00000001$; and Payoff Matrices Small High Bear and Small High Bull, see Appendix C.5 and C.6. (a) Trapping Zone Orbit (b) Attractor Arc (c) Labor Strategy Oscillation Over Three Market Periods. In (a), the trapping zone orbit is plotted in yellow, and attractor positions at $n=0$ and $n=1$ are represented in orange and blue respectively. In (b), we plot a reference attractor arc in purple with attractor positions when $n=0$, $n=0.25$, $n=0.5$, $n=0.75$ and $n=1$. (c) visualizes the fluctuation in firm and laborer preferences for gig strategies over three market periods.}
\end{figure*}

When the market shifts to bull market conditions, high-skill firms increase their preference for gig work while laborers increase their preference for employee roles, see Figure 11,12. High-skill employees have the opportunity to accrue additional compensation with a carried interest bonus, a share of profits that depend on the company's performance. As optimistic market conditions can serve as an appropriate proxy for increasing company revenues, the value of this carried interest bonus is highest during a bull market. Therefore, high-skill laborers have an increased incentive to participate in the labor market as an employee during a bull market to capture this bonus. An encouraging market outlook can also champion high-skill firms to pursue a broader range of new programs and products in different industry domains. For firms, gig labor provides a flexible on-demand pool of diverse skills that can accommodate these new risk-seeking programs. Simultaneously, employee talent acquisition and retention costs increase as laborers gain access to additional alternative work options during bull market conditions, further incentivizing firms to hire gig workers over employees during a bull market.

\begin{figure*}[h!]
\makebox[\textwidth]{
\centering
\includegraphics[width=13cm]{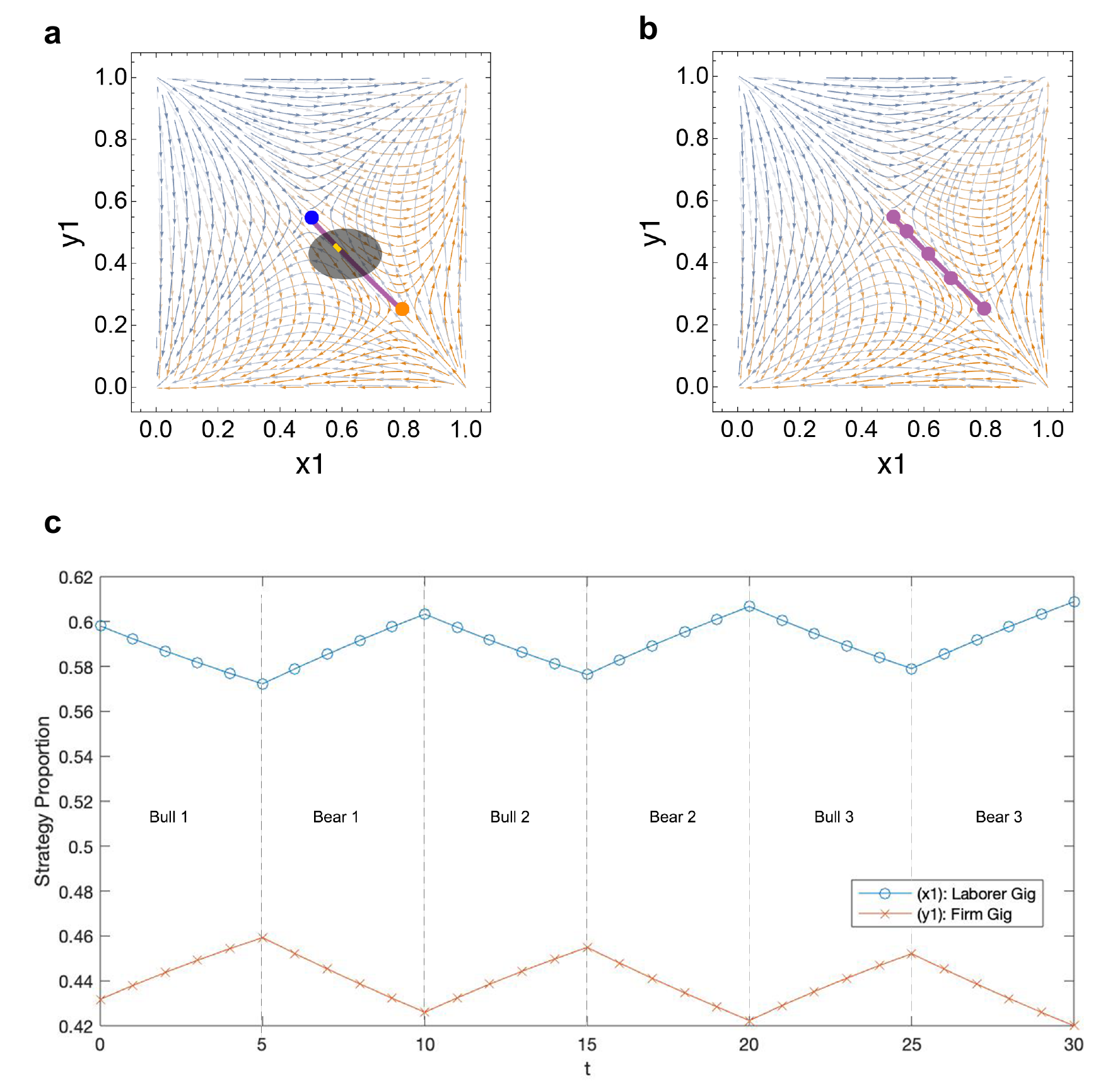}
}
 \vspace*{2mm}\caption{Evolution of Strategy Densities for Large High-Skill Firm with Initial Conditions $(0.5973, 0.4302)$, the attractor position at $n=0.5$, an approximation for a point in the trapping zone; $\omega=0.0000000002$; and Payoff Matrices Large High Bear and Large High Bull, see Appendix C.7 and C.8. (a) Trapping Zone Orbit (b) Attractor Arc (c) Labor Strategy Oscillation Over Three Market Periods. In (a), the trapping zone orbit is plotted in yellow, and attractor positions at $n=0$ and $n=1$ are represented in orange and blue respectively. In (b), we plot a reference attractor arc in purple with attractor positions when $n=0$, $n=0.25$, $n=0.5$, $n=0.75$ and $n=1$. (c) visualizes the fluctuation in firm and laborer preferences for gig strategies over three market periods.}
\end{figure*}

\subsection{Technology Influences on Firm and Laborer Gig Preference}
After we demonstrate that the system oscillating within the trapping zone reflects observable fluctuations in gig strategy densities across market conditions, we proceed to explore the role of technology in the gig economy. 

In this theoretical extension, we introduce a framework that demonstrates how technology influences labor payoffs and the growth of the gig economy. To begin, we analyze the nature in which evolving payoffs, $\dot{A}, \dot{B} \neq0$, shift the position of the attractor arc. We use the theoretical GameState pair, see Figure 1, as our reference payoff matrix pair. Let us assume that the reference payoff matrix pair represents present-day payoffs. As shown below, the reference attractor arc is rendered in yellow. 

To demonstrate the position of the attractor arc when gig strategies offer high payoffs, we add 10 to the reference payoff matrix pair for all matching gig strategies. The attractor arc for high gig payoffs is represented in blue, see Figure 15a. 
\vspace{-3mm}
\begin{figure*}[h!]
    \centering
    \makebox[\textwidth]{
    \centering
    \includegraphics[width=17cm]{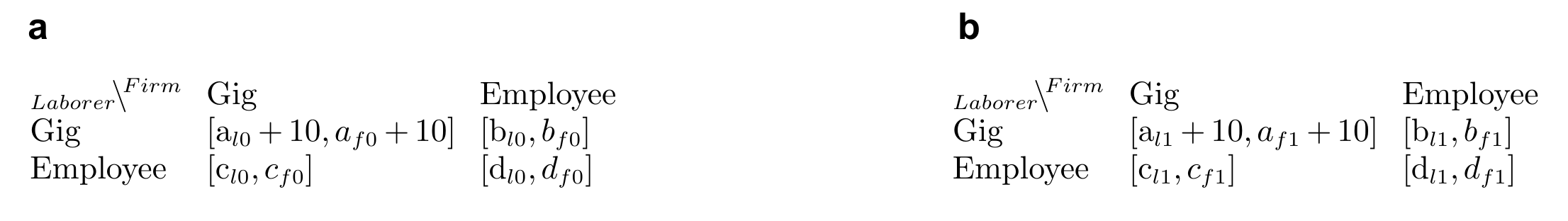}
    }
    \caption{High Gig Payoff, Matrix Operation. (a) High Gig Payoff, $n=0$ (b) High Gig Payoff, $n=1$}
\end{figure*}

To demonstrate the position of the attractor arc when employee strategies offer high payoffs, we add 10 to the reference payoff matrix pair for all matching employee strategies. The attractor arc for high employee payoffs is illustrated in red, see Figure 15b. 
\vspace{-3mm}
\begin{figure*}[h!]
    \centering
    \makebox[\textwidth]{
    \centering
    \includegraphics[width=17cm]{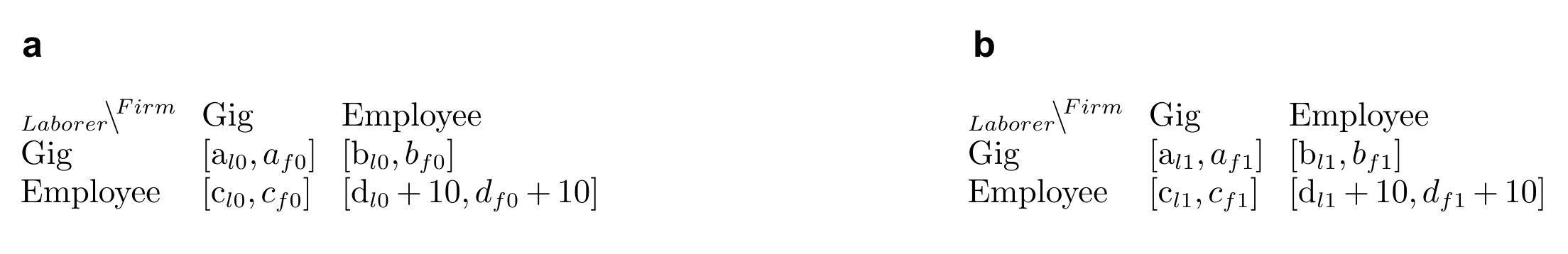}
    }
   \caption{High Employee Payoff, Matrix Operation. (a) High Employee Payoff, $n=0$ (b) High Employee Payoff, $n=1$}
\end{figure*}

\begin{figure*}[h!]
    \centering
    \includegraphics[width=13cm]{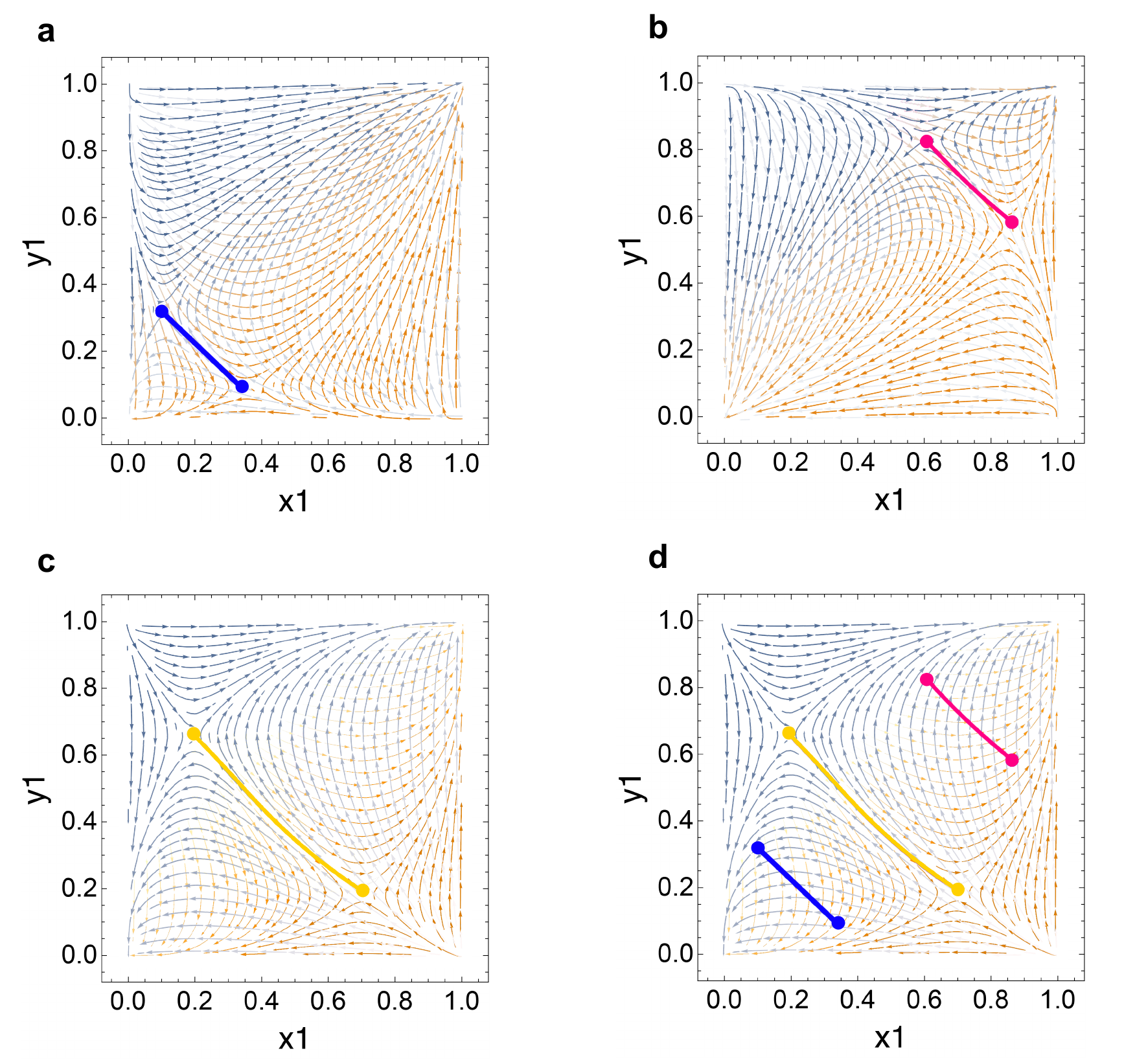}
   \vspace{2mm}\caption{Attractor Arc Drift Transformations (a) Arc transformation with High Gig Payoff Matrix Operation, see Figure 13, and Theoretical Gamestate Pair, see Figure 1 (b) Arc transformation with High Employee Payoff Matrix Operation, see Figure 14, and Theoretical Gamestate Pair, see Figure 1 (c) Reference attractor arc with Theoretical Gamestate Pair, see Figure 1 (d) Composite diagram with arcs (a), (b) and (c).}
\end{figure*}

\paragraph{Technology and the Neoteric Growth of the Gig Economy} 

In recent decades, the gig economy has ballooned from relative obscurity to more than one third of the labor market \citep{gallup2018gig}. Mapped to our model, this growth implies that the attractor arc evolved from a region near $(0,0)^*$ towards $(1,1)^*$; this is pictured as a shift from the blue arc to the yellow arc, indicating an increase in gig workers. 

The scenario with high gig payoff represents the premature gig economy, an era predating digital platforms. In this premature economy, gig positions were elite, skilled roles. An angel investor or company advisor or perhaps even a Mckinsey consultant typified the variety of early gig positions \citep{hyman2018temp}. Here, gig workers provide much higher payoffs than employees. Since payoff is determined by compensation, high payoff implies high compensation. Appropriately, when gig payoffs are very high, each company can afford to hire a small amount of these elite gig workers. This explains why the high gig payoff attractor arc is near $(0,0)^*$, indicating a labor composition consisting of few gig workers.

More recently, technology has enabled low-skill workers to sustainably participate in the gig economy. Examples of such technologies include ride-sharing and last-mile delivery apps such as Uber, Lyft and Doordash as well as freelancing websites such as Upwork, all of which introduce mostly low-skill, low-payoff workers to the gig economy. As low-skill gig workers such as Uber drivers flooded the gig economy, gig payoffs decreased relative to employee payoffs. This is consistent with our model as the attractor arc shifts towards $(1,1)^*$ from the blue to yellow arc during this development, reflecting the neoteric growth of the modern gig economy.

\paragraph{Technological Implications on the Future of the Gig Economy}
Notional future growth of the gig economy is represented by the evolution from the yellow arc to the red arc. Per our model, as employee payoffs increase relative to gig payoffs, the attractor arc nears $(1,1)^*$; this implies that the labor market consists of mostly gig workers and few employees. Some ride-sharing firms may already example such distinct gig-employee bifurcation consistent with an arc positioned near $(1,1)^*$. For instance, Uber's personnel consists of many low payoff gig drivers, and relatively few high payoff engineers, managers and executives. Such a distribution is reflected in our model as we observe an attractor arc position higher up on the $y_1$ axis for low-skill firms, implying a workforce with a higher density of gig laborers.

There are cogent reasons to believe that the gig economy might either decrease or increase in size, a tension we aim to inform. We offer model-informed explanations that acknowledge the two competing logics. In order for the gig economy to continue growing, employee payoffs must increase relative to gig payoffs. Such a development implies that high skill work must advantage gig roles such that executives, the highest paid individuals, are the only employees remaining in an enterprise. In the current enterprise structure, there are numerous obstacles facing such a workforce transformation. While low-skill firms compete on pricing, high skill firms compete on talent. Thus far, most gig-dominant firms are low-skill firms such as Uber and Lyft which leverage commodity skill workers to operate their services. On the other hand, the notion of ubiquitous high-skill gig work faces the legal and strategic complication of trade secrets, non disclosure agreements, non-competes, and other intellectual property complexities. Further, there is growing consensus that artificially intelligent machines will replace many processes currently fulfilled by commodity-skill human operators. Resultantly, low-skill gig workers such as Uber drivers will be displaced as a part of this technological transformation, signaling a future contraction in the present day low-skill dominant gig economy. The question is whether these displaced workers will find new roles as employees or gig workers. 

There are also compelling reasons to believe that the gig economy will continue growing. Researchers have conjectured that workers displaced by AI technologies will find roles in which they supervise machines and fulfill other more creative responsibilities \citep{agrawal2018prediction}. Creative roles are a suitable fit for the gig economy as these positions champion worker flexibility. While ride-share companies like Uber and Lyft may decrease their gig application, the freelancing cohort of the gig economy may potentially continue growing. Further, the future may entail a re-constitution of enterprise with pioneering frontier technologies, decentralization and remote work arrangements. A reconstitution of policy structures can also play a role in the regulation and protection of trade secrets, all of which may support adoption of ubiquitous high-skill gig work. 

The work and enterprise structures of the future depend on a dizzying constellation of cultural and technological developments, rendering it difficult to speculate the future direction of the gig economy. While we address the competing logics, we do not state a specific preference for future gig economy growth or contraction. We hope that our model extension can inform the discussion by providing a new payoff framework that can be applied when thinking about technology's role in the growth of the gig economy. 

\subsection{Policy Influences on Firm and Laborer Gig Preference}
Using similar approach, we also explore policy influences on labor strategies by applying an evolving-payoff framework. While the gig economy has been viewed as beneficially transformative to some, others share a more precarious disposition regarding its economic imbalances. For researchers, policy makers and industrialists alike, there exists a tension as to whether or how to regulate the gig economy. In the context of the labor market, policy behaves as a mechanism that can transfer risk and economic burdens between firms and laborers \citep{isaac2014disruptive, johnston2018organizing, todoli2017end}.

During periods of lenient policy regulation, firms can take advantage of regulatory ambiguity and exploit gig workers. On the other hand, gig laborers are unprotected and must tolerate firm expectations. To model the payoff during a period of lenient policy ordinance, we subtract 3 from the laborer's gig payoff and add 3 to the firm's gig payoff. The attractor arc for lenient policy ordinance is represented in red, see Figure 18a. 

\begin{figure*}[h!]
    \centering
   \makebox[\textwidth]{
    \centering
    \includegraphics[width=17cm]{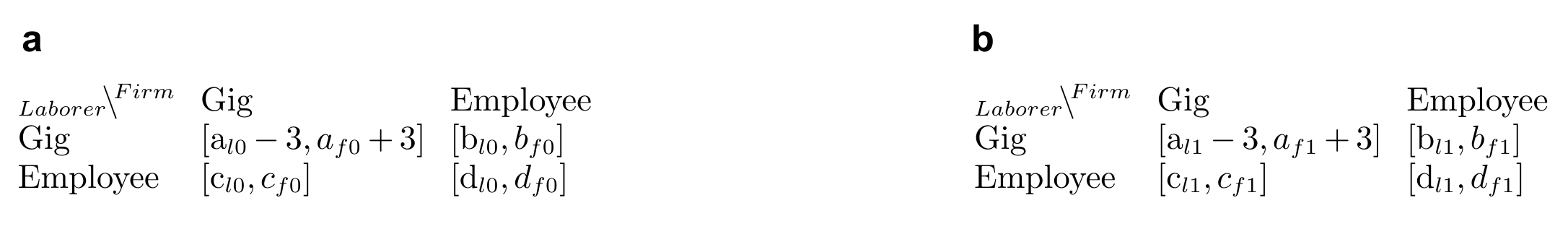}
    }
   \caption{Lenient Policy, Matrix Operation. (a) Lenient Ordinance, $n=0$ (b) Lenient Ordinance, $n=1$}
\end{figure*}

In the course of strict policy enactment, governments demand firms to more closely classify gig workers as employees. For instance, a government may mandate that firms provide benefits and additional protections to gig laborers. Accordingly, gig workers benefit as they receive additional worker protections and increased welfare. To model the payoff during a period of strict policy ordinance, we add 3 from the laborer's gig payoff and subtract 3 to the firm's gig payoff. The attractor arc for strict policy ordinance is represented in blue, see Figure 18b. 

\begin{figure*}[h!]
    \centering
    \makebox[\textwidth]{
    \centering
    \includegraphics[width=17cm]{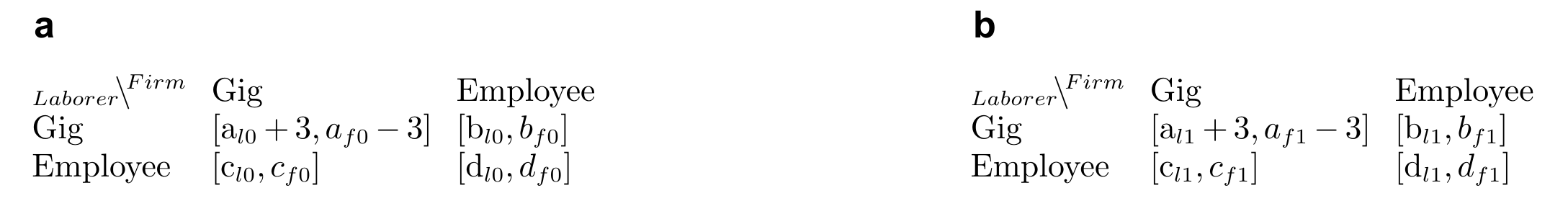}
    }
    \caption{Strict Policy, Matrix Operation. (a) Strict Ordinance, $n=0$ (b) Strict Ordinance, $n=1$}
\end{figure*}

\begin{figure*}[h!]
    \centering
    \includegraphics[width=13cm]{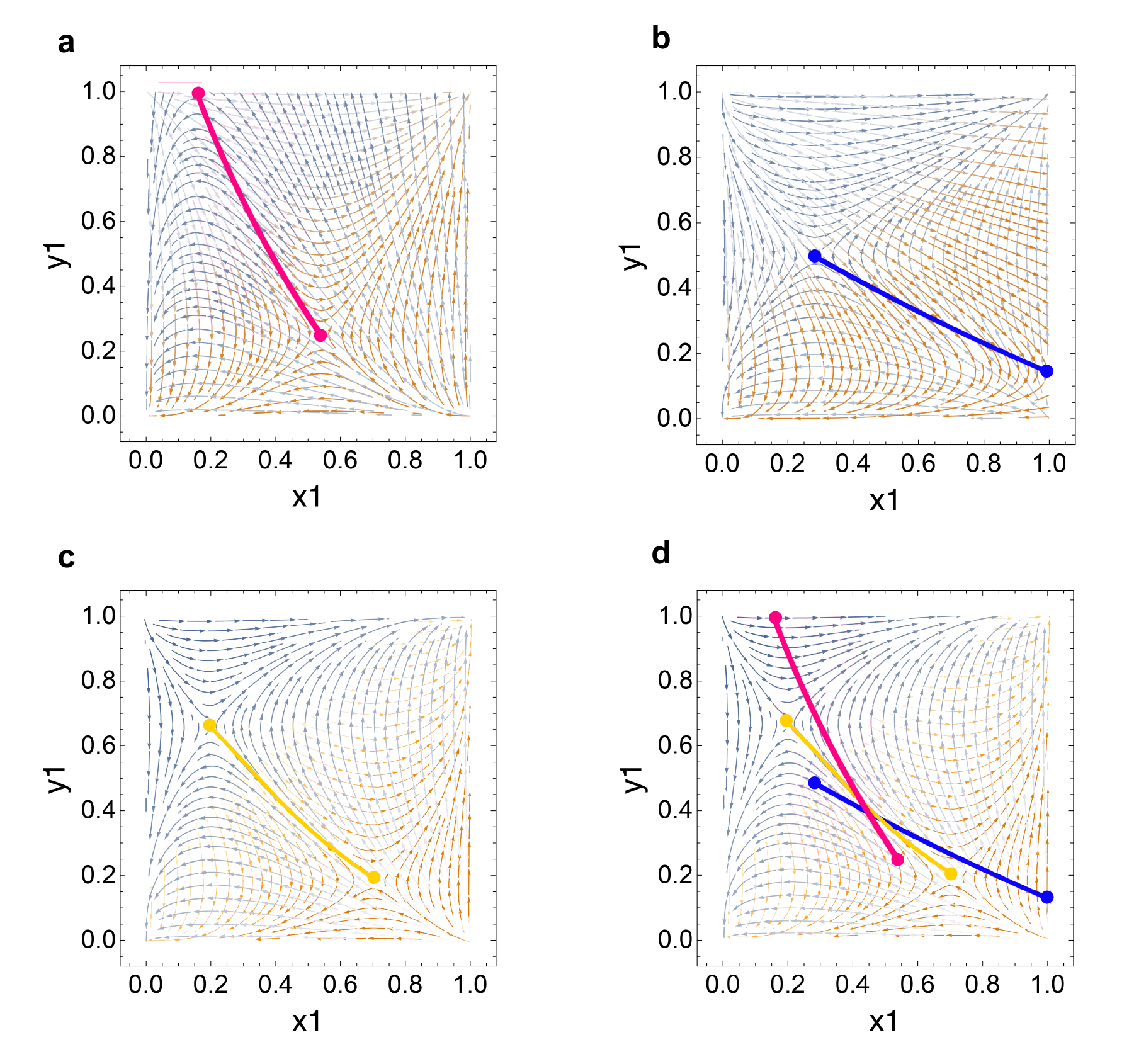}

   \caption{Attractor Arc Drift Transformations. (a) Arc transformation with Lenient Policy Matrix Operation, see Figure 16, and Theoretical Gamestate Payoff Pair, see Figure 1 (b) Arc transformation with Strict Policy Matrix Operation, see Figure 17, and Theoretical Gamestate Payoff Pair, see Figure 1 (c) Reference attractor arc with Theoretical Gamestate Payoff Pair, see Figure 1 (d) Composite diagram with arcs (a), (b) and (c).}
\end{figure*}

\paragraph{The Impact of Regulation on Labor Strategy Sensitivities.} While the position of the attractor arc is a suitable proxy for the position of the trapping zone, arc orientation does not always represent the orientation of the trapping zone. As the slope of the arc increases and the arc becomes more vertical, the slope of the trapping zone decreases and becomes more horizontal. Conversely, as the slope of the arc decreases and the arc becomes more horizontal, the slope of the trapping zone increases and becomes more vertical. This concept is visually represented in Figure 19. On one hand, when the attractor arc becomes more vertical, laborers experience an increased sensitivity between employee and gig strategies across market cycles while firms experience a decreased sensitivity; this can be understood as oscillators in the trapping zone become elongated on the $x_1$ axis and shortened on the $y_1$ axis. On the other hand, when the attractor arc becomes more horizontal, laborers experience an decreased sensitivity between employee and gig strategies across market cycles while firms experience an increased sensitivity; this can be understood as oscillators in the trapping zone become shortened on the $x_1$ axis and elongated on the $y_1$ axis. We define an oscillator as an evolutionary orbit for the system across market cycles. We define sensitivity as the distinction between and preference for gig or employee strategies across market conditions. 

\begin{figure*}[h!]
    \centering
    \makebox[\textwidth]{
    \centering
    \includegraphics[width=13cm]{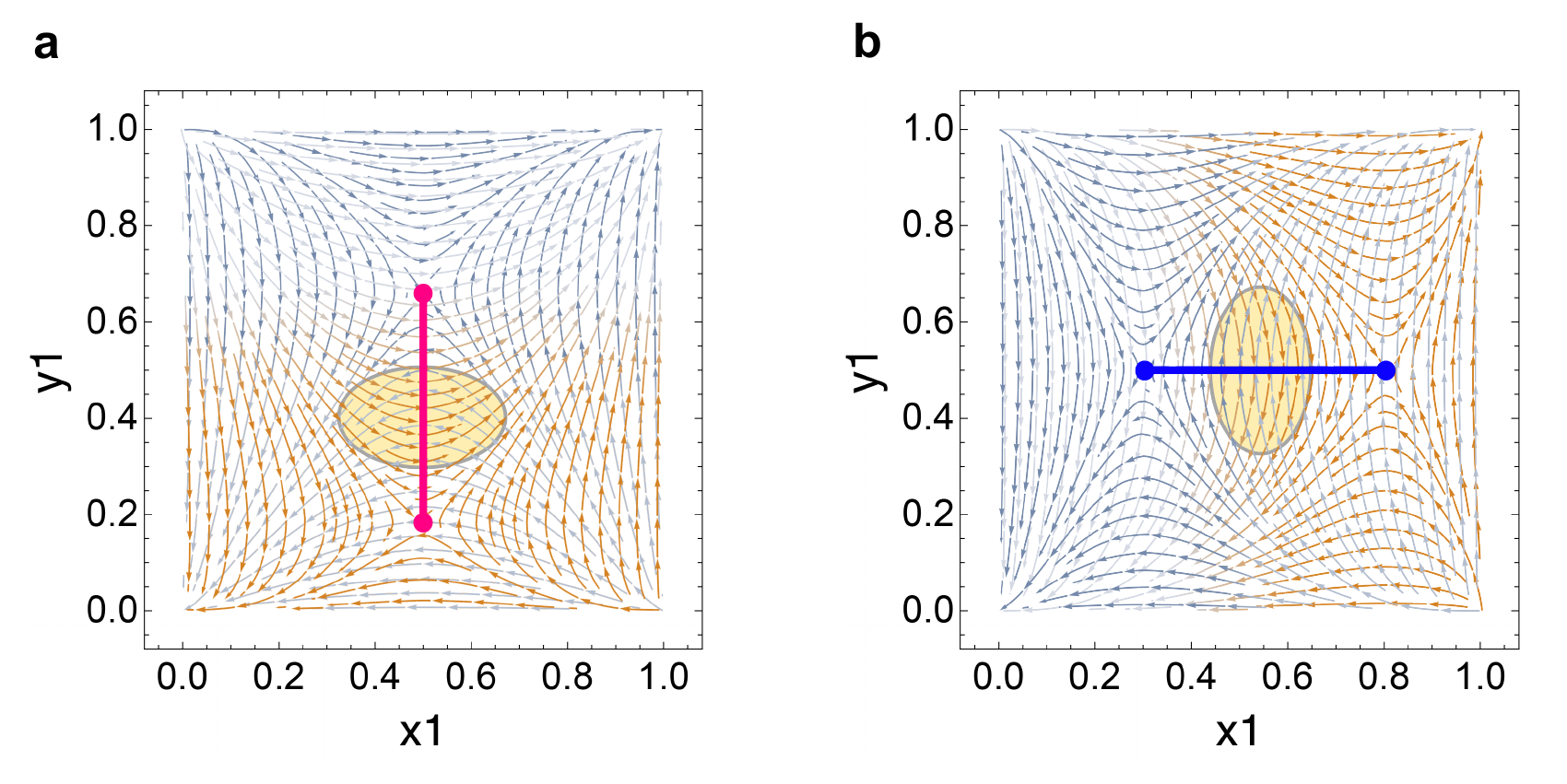}
    }
    \caption{Vertical and Horizontal Attractor Arc and Trapping Zone Slopes. (a) Attractor arc using theoretical payoff pair, see Appendix C.9. When the attractor arc is oriented vertically, the slope of the trapping zone becomes horizontal and perpendicular to the arc. (b) Attractor arc using theoretical payoff pair, see Appendix C.10. When the attractor arc is oriented horizontally, the slope of the trapping zone becomes vertical and perpendicular to the arc. The opaque yellow ellipse is a background element to indicate the trapping zone. The evolutionary trajectories in both (a) and (b) trapping zones are orthogonal to their respective arcs. }
\end{figure*}

An interval of lenient policy will drive the attractor arc to increase in slope and become more vertical while strict policy will drive the arc to decrease in slope and become more horizontal. For our demonstrations, we will use our vertical attractor arc as an extreme example of lenient policy and our horizontal attractor arc as an extreme example of strict policy.

During a period of strict regulatory ordinance, the firm must pay the gig worker increased compensation, even though the gig worker provides the same quality of work as before. Therefore, the firm experiences an increased sensitivity and larger distinction between gig workers and employees. If we consider our horizontal attractor arc to be an extreme example of strict policy, we see that the $y_1$ trapping zone span is elongated while the $x_1$ trapping zone span is shortened. The longer $y_1$ trapping zone span exhibits the firm's increased sensitivity to worker type and a greater distinction between hiring gig workers or employees. For laborers, the shorter $x_1$ span signifies a decreased sensitivity for participating as a gig worker or an employee; this is a logical transformation, as strict policy mandates greater equality in the treatment of gig workers and employees, forming a strengthened gig-employee resemblance. 

Conversely, in a period of lenient policy denoted by the red arc, we find that the $x_1$ trapping zone span is elongated while the $y_1$ trapping zone span is shortened. Considering the lack of gig worker protections during intervals of lenient policy, it is sensible that gig workers experience increased sensitivity between worker categories without regulation, as there is greater distinction between working as an employee or a gig worker. On the other hand, firms experience a decreased sensitivity for worker type as they can take advantage of regulatory ambiguity to maximize operational efficiency. 

In this theoretical extension, we assume that policy behaves as a mechanism that can shift economic burdens, represented through payoffs, between firms and laborers. We propose an evolving-payoff framework to model the impact of policy regulations on firm and worker labor strategies. Our findings inform existing literature and scholarship by demonstrating how policy transfers payoff utility and alters firm and laborer sensitivities for different labor strategies.

\section{Discussion}

The emergence of the modern gig economy introduces a new set of employment considerations for firms and laborers. Among manifold regards, firms must elect between hiring a gig worker or an employee while balancing labor costs with product quality and worker reliability. When deciding to participate in the gig economy, laborers must evaluate autonomy at the expense of financial stability and labor protections conferred with employee status. In practice, these elements of employment incentives and deterrents can be modeled with strategy-dependent payoffs, presenting a suitable opportunity for a game theoretical exploration. Influenced by several macroeconomic forces, these employment incentives are shaped by the nexus between dynamic market, technology and policy developments. On one hand, a bear market can discount worker-autonomy and accessible service demand from consumers. On the other, a bull market can enable workers to engage in a broader scope of alternative engagements and earn additional bonuses. Indeed, high and low skill laborers are impacted differently and have idiosyncratic susceptibilities to market changes. Regarding regulatory structures, policy behaves as a mechanism that transfers economic burdens between firms and laborers. For researchers and policy markers alike, there remains an unanswered question as to whether or how to regulate the gig economy. Adjacently, advancements in technology - in particular, digital platforms - have often been attributed as catalyst of growth for the modern gig economy. Contrarily, other technologies such as AI may implicate a future contraction of the servicing gig economy. Consolidating a multitude of micro and macro determinants, we explore how the composition of firm and laborer strategies for gig or employee labor evolve under different market conditions, regulatory ordinances and technological expansions.

In our work, we apply a game theoretical approach to study the evolution of strategy densities in firm and laborer populations, recasting employment incentives into strategy-dependent payoffs and fluctuating market conditions into an evolving environment variable. Formally, we extend the replicator equation to model oscillating dynamics in two-player asymmetric bi-matrix games with a time-evolving environment. While classical game theory centers on stable equilibrium solutions, we demonstrate a pseudo-stable state in which the system oscillates in a trapping zone orbit as a result of dynamic payoffs governed by an evolving environment. We extend our model to exhibit how changes in payoffs can transform the orientation and position of the system's oscillatory orbit, concepts we refer to as arc drift and arc tilt. Applying these concepts to our study of the gig economy, we demonstrate how technology and policy can implicate arc drift and tilt. 

Here we present four noteworthy contributions to existing scholarship on the gig economy and evolutionary dynamics. First, we extend the replicator equation to a new form of game, oscillating replicator dynamics with attractor arcs, introducing concepts of the attractor arc, driven oscillation, trapping zone and escape. We extensively study the behavior of a pseudo-stable equilibrium which is governed by an evolving environment variable. We detail this pseudo-stable equilibrium with the notion of a trapping zone. In our model, we suggest that the system will forever remain in this pseudo-stable state and never escape to an ESS. Under this logic, we are able to demonstrate how changing payoffs result in a variety of attractor arc transformations, presenting a novel analytical approach for evolutionary game theory. 

Second, we discover that market conditions implicate different evolutionary patterns for strategy densities in high and low skill firms. Low-skill firms and laborers decrease preference for gig work in favor of employee strategies during bear markets and both populations favor gig strategies during bull markets. While low-skill firms and laborers demonstrate a matching oscillatory behavior (when firm preference for gig work increases, laborer preference for gig strategies also increases), high skill firms and laborers exhibit a mismatching oscillatory behavior. During a bear market, high skill firms decrease their preference for gig work while laborers increase their participation in the gig economy. Under bull market conditions, high skill firms increase their preference for gig work, while laborers decrease gig participation in favor of employee roles. We explain this behavioral polarity between the high and low skill work-forces through their differing sensitivities to market-driven consumer demand, operational requirements and financial incentives among other considerations. 

Third, we propose a payoff framework to analyze the role of technology in the growth of the gig economy, informing tensions regarding the future of this new employment category. By exploring the nature of attractor arc drift, we establish payoff operations that imply the growth or contraction of the gig economy. Consistent with historical observations, our model suggests that the early gig economy consisted of elite, specific skilled roles with high payoffs such as a management consultant or company advisor. We provide analysis that suggests technology, namely digital platforms, enabled low skill workers to sustainably participate in the gig economy, resulting in its neoteric rise. In our theoretical extension, we offer arguments that suggest the gig economy may either continue to grow or contract in the future. The direction of future gig economy growth depends on various technological developments and a potential future re-constitution of work and enterprise. 
 
Fourth, we explore regulatory implications within the gig economy, demonstrating how policy acts as a mechanism to transfer risk and economic burden between firms and laborers. In our model, we investigate the impact of shifting payoff utility between firms and laborers. We find that intervals of lenient and strict regulatory ordinances alter firm and worker sensitivities to different labor strategies.

This work is founded on assumptions contingent on a number of limitations. We present our model's constraints, mapping out directions with promising opportunity for future research. 

Our payoff generation methodology is founded on several inferences. First, we greatly simplify the enterprise landscape by creating discrete buckets for firms and contracts. Second, our payoff generation methodology considers only a small selection of possible employment incentives (compensation, reliability, flexibility, talent retention and potential alternative engagements). Other employment considerations for laborers not included in our model are worker status, career mobility, stress, and isolation among others. For firms, the model can be extended to consider enterprise-scalability, diversity, culture and taxes. Third, we are responsible for quantifying discount weights for each payoff coefficient. We apply reasonable assumptions to generate our payoffs and present a theoretical framework from which further empirical extensions can be investigated. 

As discussed, our evolutionary model presents a pseudo-stable equilibrium conditional on the relationship between selection intensity, $\omega$, and the rate of environment evolution, $\Dot{n}$. We demonstrate oscillation in the trapping zone with a non-continuous $\Dot{n}$ function and a small $\omega$ value. Further research can be conducted to explore specific escape velocities, continuous $\Dot{n}$ functions and additional estimates for points in the trapping zone to be used as initial conditions. Moreover, we also hypothesize that there exist regions in some systems wherein which infinite oscillation in a trapping zone is possible; research on the alignment of attractor arcs and system symmetries may elucidate on this hypothesis.

Although we introduce the concept of escape, we do not throughly investigate this potential scenario. While we maintain that the system likely will never escape to an ESS, there is a potential area of research on scenarios in which the system \emph{escapes} but is \emph{recaptured}. It is plausible, for instance, that the model self corrects in order to always trap the system such that it never fully escapes to an ESS. Take for example the 2020 global COVID-19 pandemic, a widely disruptive event that may accelerate $\omega$ or $\Dot{n}$ such that the system quickly reaches escape boundary and escapes towards an ESS. Escape may reflect a rapid decline in gig strategy density as demand for consumer services curtails during the national shutdown of non-essential businesses. Government response such as the passing of a multi-trillion dollar stimulus may be a system correction that may shift the arc in an attempt to re-capture and trap the escaped system. In this hypothetical example, escape is complementary to our thesis that the system will never escape to an ESS; rather than implicating the system to escape to an ESS, a perturbation that causes escape may be re-captured by means of an external force or action such as governmental intervention. Finally, it is our hope that scholarship on the gig economy can extend to study adjacent topics of education and economic mobility. Perhaps, the rise of distributed and widely-accessible education resources paired with a reconstitution of work and enterprise will establish the future gig economy as a means of economic mobility. 

In conclusion, we propose a model that incorporates a co-evolving treble of macro forces -- markets, technology and policy -- and demonstrate their respective influences on labor strategies in the gig economy. We demonstrate how technology is a driver of change in the labor economy and how policy is integral to the sustainability of new systems and the protection of involved parties. The primary goals of this paper are to further comprehension of micro and macro influences on firm and laborer incentives for gig adoption. We provide researchers, policy makers and industrialists alike with a novel evolutionary model and payoff framework approach for better understanding firm and laborer behaviors in the gig economy. 

\section*{Acknowledgements}
We are grateful for support from the Bill \& Melinda Gates Foundation (award no. OPP1217336), the NIH COBRE Program (grant no.1P20GM130454), the Neukom CompX Faculty Grant, the Dartmouth Faculty Startup Fund, and the Walter \& Constance Burke Research Initiation Award.

\newpage

\appendix

\section{Payoff Generation}
In this next section, we generate utility payoffs for firms and laborers across firm category and market condition. Our model evaluates four categories of firms; the categories are defined as \emph{Small Low-Skill Firm}, \emph{Small High-Skill Firm}, \emph{Large Low-Skill Firm}, and \emph{Large High-Skill Firm}. For each firm category, a pair of payoff matrices are generated to represent the firm in bear and bull market conditions. 

\subsection{GameStates, Contracts and Payoff Coefficients}

 A firm's operation consists of numerous processes or tasks that must be effectuated within a time interval in order for the firm to maintain operation and stay competitive. We simplify the myriad of possible processes by representing these tasks with four contract categories; the categories are \emph{Short-Term Low-Skill Contract}, \emph{Short-Term High-Skill Contract}, \emph{Long-Term Low-Skill Contract}, and \emph{Long-Term High-Skill Contract}. Indeed, each firm's operational need translates into a discrete distribution involving these four contract categories; this is the firm's labor demand. 
 
 To fulfill each individual contract, a firm can choose to hire either a gig worker or an employee. Conversely, a laborer can participate as either a gig worker or an employee in competition of a contract. Accordingly, payoffs are assigned to each firm and laborer strategy to model the efficacy of the strategy.

A firm's labor demand distribution is determined by market condition (bear or bull), firm size (small or large), and firm skill-set (low or high skill). The eight combinations of the aforementioned three constituents (market condition, firm size and firm skill-set) constitute the eight discrete GameStates; a GameState represents one of the four firm categories in one of the two market conditions. For example, the GameState \textit{Small Low Bear} denotes a small low-skill firm in a bear market. 

In addition to discrete contract distributions, each GameState will have four coefficients for flexibility, reliability, talent retention and potential alternatives. These coefficients represent the importance of additional employment incentives for firms and laborers beyond compensation or cost of labor. Each payoff coefficient instantiates the respective weight or importance of flexibility, reliability, talent retention and potential alternatives in each GameState.  

The flexibility payoff coefficient weighs the importance of flexible labor for the firm. The reliability payoff coefficient accounts for the significance of labor quality and worker reliability for the firm. The talent retention payoff coefficient represents the firm's cost of obtaining and retaining labor talent. Finally, the potential alternatives coefficient denotes the laborer's potential utility from participating in alternative activities outside the contract, serving as a proxy for laborer flexibility. In sum, payoff coefficients are additional employment considerations for firms and laborers.

For each GameState constituent (Firm size (small or large) denoted by $F_{size}$, Firm skill set (high or low-skill) denoted by $F_{skill}$, and market denoted by $M$ (bear or bull)), we assume a contributing weight for each payoff coefficient. For each GameState, i.e., Small Low Bear, payoff coefficients are calculated by taking the summation of applicable weights. To illustrate an example, the potential alternatives coefficient $\Pi_P$ is determined by firm size, firm skill and market condition. In the $Small Bear Low$ Gamestate, the potential alternatives coefficient is a combination of potential alternatives weights assigned by $F_{size}:Small$, $F_{skill}:Low$ and $M:Bear$. 
\\\\
$\Pi_P(F_{size}, F_{skill}, M) = \Pi_P(F_{size}:Small) + \Pi_P(F_{skill}:Low) + \Pi_P(M:Bear)$
\\\\
GameState constituents also influence how a firm partitions a labor budget across different types of contracts. We apply a similar operation to determine the proportion of high and low-skill contracts and the proportion of long and short contracts. 

\subsubsection{GameState Constituent Assumptions}
In this section, we specify the contract proportions and payoff weights for each GameState constituent. $\gamma_{(S)}$ denotes the weighted fraction of short contracts, $\gamma_{(L)}$ denotes the weighted fraction of long contracts, $\Pi_F$ denotes flexibility, $\Pi_R$ denotes reliability, $\Pi_T$ denotes talent retention, $\Pi_P$ denotes potential alternatives, $\mu_{(Lo)}$ denotes the fraction of low skill contracts and $\mu_{(Hi)}$ denotes the fraction of high skill contracts. 

\begin{flushleft}
\textbf{$M$ Bear: ($\gamma_{(S)}$: +$\frac{3}{10}$, $\gamma_{(L)}$: +$\frac{2}{10}$, $\Pi_F$: $+1$, $\Pi_R$: $+5$, $\Pi_T$: $0$, $\Pi_P$: $0$) }\par
\end{flushleft}
    We assume that bear market conditions incentivize firms to become more risk averse in their long term strategies and therefore spend more conservatively on long term projects and risky innovation. Simultaneously, we reason that firms will focus resources on flexible short term strategies that allow for quick adaptability to unfavorable developments. Accordingly, we assign a bear market influence of +$\frac{3}{10}$ and +$\frac{2}{10}$ for a firm's demand for short and long term contracts respectively. 
    
    For payoff coefficients, we assign a weight of +1 for \emph{flexibility} to account for fluctuating business needs and strategy pivots from immediate market stressors and a weight of +5 for \emph{reliability} by reason of a lower threshold for and higher cost of error in a bear market. Weights for \emph{talent retention} and \emph{potential alternatives} are unaffected by bear market conditions.

\begin{flushleft}
\textbf{$M$ Bull: ($\gamma_{(S)}$: +$\frac{2}{10}$, $\gamma_{(L)}$: +$\frac{3}{10}$, $\Pi_F$: $0$, $\Pi_R$: $0$, $\Pi_T$: $+7$, $\Pi_P$: $+5$) }\par
\end{flushleft}

    We expect that bull market conditions will incentivize firms to become more risk seeking in their long term strategies and therefore spend more aggressively on risk-seeking innovation bets and long term strategies. As a result of optimistic market conditions, firms can plan ahead with more foresight and integrate short term requirements into longer term programs. Accordingly, we assign a bull market weight of +$\frac{2}{10}$ and +$\frac{3}{10}$ for a firm's demand for short and long term contracts respectively. 

    A bull market can serve as a proxy for low unemployment rates. For payoff coefficients, we assign a weight of +7 for \emph{talent retention} as unemployment rates are low, and there are more opportunities for workers to pursue, thereby increasing the cost of talent acquisition and retention for firms. Further, we assign a weight of +5 for \emph{potential alternatives} as laborers have access to pursue a broader range of alternative engagements in a bull market. Weights for \emph{flexibility} and \emph{reliability} are unaffected by bull market conditions.

\begin{flushleft}
\textbf{$F_{size}$ Small: ($\gamma_{(S)}$: +$\frac{4}{10}$, $\gamma_{(L)}$: +$\frac{1}{10}$, $\Pi_F$: $+10$, $\Pi_R$: $+2$, $\Pi_T$: $0$, $\Pi_P$: $0$) }\par
\end{flushleft}

	We conjecture that small firms behave more dynamically in the short term due to increased pivots as they develop product-market fit and build out their operations. Resultantly, we expect small firms to focus on short term strategy in order to accommodate changing business requirements and concentrate resources on immediate operational needs. Accordingly, we assign a small firm influence of +$\frac{4}{10}$ and +$\frac{1}{10}$ for a firm's demand for short and long term contracts respectively.

    For payoff coefficients, we assign a weight of +10 for \emph{flexibility} by logic of increased agility, fluctuating business needs and strategy pivots attributable to small businesses. A weight of +2 is designated to \emph{reliability} because each individual's contribution and responsibility is more substantial in a smaller team, thereby increasing the impact of error for each individual. Weights for \emph{talent retention} and \emph{potential alternatives} are unaffected by small firm size.
    
\begin{flushleft}
\textbf{$F_{size}$ Large: ($\gamma_{(S)}$: +$\frac{1}{10}$, $\gamma_{(L)}$: +$\frac{4}{10}$, $\Pi_F$: $0$, $\Pi_R$: $0$, $\Pi_T$: $0$, $\Pi_P$: $0$) }\par
\end{flushleft}

    	We reason that large firms behave less dynamically in the short term as a result of established, sustainable business models and a lower likelihood of pivoting at size. Accordingly, we assign a large firm influence of +$\frac{1}{10}$ and +$\frac{4}{10}$ for a firm's demand for short and long term contracts respectively.
	
	    For payoff coefficients, weights for \emph{flexibility}, \emph{reliability}, \emph{talent retention} and \emph{potential alternatives} are unaffected by large firm size.
\begin{flushleft}
\textbf{$F_{skill}$ Low: ($\mu_{(Hi)}$: $\frac{2}{10}$, $\mu_{(Lo)}$: $\frac{8}{10}$, $\Pi_F$: $0$, $\Pi_R$: $0$, $\Pi_T$: $0$, $\Pi_P$: $0$)
 }\par
\end{flushleft}

	We presume that low-skill firms maintain an operational demand distribution split between 20 percent high skill and 80 percent low skill contracts. We expect that a firm's required skill set does not impact the demand distribution of short and long term tasks.
	
   For payoff coefficients, weights for \emph{flexibility}, \emph{reliability}, \emph{talent retention} and \emph{potential alternatives} are unaffected by large firm size.
	
\begin{flushleft}
\textbf{$F_{skill}$ High: ($\mu_{(Hi)}$: $\frac{8}{10}$, $\mu_{(Lo)}$: $\frac{2}{10}$, $\Pi_F$: $0$, $\Pi_R$: $+10$, $\Pi_T$: $+3$, $\Pi_P$: $+5$ ) }\par
\end{flushleft}

	We presume that high-skill firms maintain an operational demand distribution split between 80 percent high skill and 20 percent low skill contracts. We assume that a firm's required skill set does not impact the demand distribution of short and long term tasks.
	
   For payoff coefficients, weights for \emph{flexibility}, \emph{reliability}, \emph{talent retention} and \emph{potential alternatives} are unaffected by large firm size.

\subsubsection{Compounded GameState Constituent Assumptions}

\begin{flushleft}
\textbf{$Firm_{size}$ Large $and$ $Firm_{skill}$ High: ($\Pi_F$: $+5$, $\Pi_R$: $-3$)}
\end{flushleft}

Large high-skill firms such as Microsoft are consistently pursuing a breadth of projects ranging across industries. We apply a $flexibility$ weight of +5 to account for the variety of skills required to accommodate this wide-ranging horizon of programs and projects. Although high-skill labor warrants an increased sensitivity to labor reliability, we assign large high skill firms a $reliability$ weight of -3 as large team size reduces the average impact of error for each worker.

\begin{flushleft}
\textbf{$Firm_{size}$ Large $and$ $M$ Bull: ($\Pi_T$: $-5$)}
\end{flushleft}

For large firms in a bull market, we assign \emph{talent retention} a weight of -5. We reason that large companies can leverage corporate brand names to attract a larger and more consistent pool of applicants.

\begin{flushleft}
\textbf{$Firm_{size}$ Small $and$ $Firm_{Skill}$ Low: ($\Pi_F$: $-7$)}
\end{flushleft}

 For small low-skill firms, we assign a $flexibility$ weight of -7. We posit that most small low-skill firms (i.e., family owned restaurants) operate static business models and experience marginal business innovation, thereby decreasing the operational demand for flexible skills.

\subsubsection{GameState Contract Demand and Payoff Coefficients}

To model operational demand, we first designate an annual labor spend to each firm category by firm size. We model payoffs in units of utility. Large and small firms are assigned $100M$ and $2M$ annual labor budgets respectively. Annual labor spend is represented with $\xi_{(Large, Small)}$. High-skill and low-skill labor, respectively cost $100K$ and $30K$ annually regardless of worker type (gig or employee). Labor cost by skill-set is denoted with $\Psi_{(Hi, Lo)}$. Short contracts span 2 weeks and long contracts span 26 week (half year) intervals.

To calculate the firm's requirement of a specific contract in a GameState, we first allocate a fraction of the annual labor spend to the type of task; each contract category captures a fraction of the firm's annual labor spend. This apportionment is determined by partitioning the labor spend according to the fraction of short or long contracts and the fraction of low or high-skill requirements. These proportions are calculated according to coefficients specified in the contract type ($ProportionLength: \gamma_{(S,L)}$) ($ProportionSkill: \mu_{(Hi,Lo)}$). Finally, we divide the partitioned budget by the cost of labor and normalize the contract count to reflect an annual interval; $\chi_{(S,L)}$ equates to 52 weeks divided by the contract duration (short or long) in weeks.  
\begin{equation}
\centering
Contract Demand_{*} = (\frac{(\xi_{(Large, Small)})(\gamma_{(S,L)})( \mu_{(Hi,Lo)})}{\Psi_{(Hi,Lo)}})(\chi_{(S,L)})
\end{equation}
\noindent \textbf{Example with Short Low Skill Contract for Small Low Bear}

\begin{equation}
\hspace{-5mm}
\centering
Demand_{shortlowskill} = (\frac{(\xi_{(Small)})(\gamma_{(S)})( \mu_{(Lo)})}{\Psi_{(Lo)}})(\chi_{(S)})= (\frac{(2000000)(\frac{7}{10})(\frac{8}{10})}{30000})(\frac{52}{2}) \approx 970
\end{equation}
With this example calculation, we see that a small low-skill firm will have an annual demand for 970 short term low skill contracts. We select the $\gamma_{(S)}$, $\chi_{(S)}$ $\mu_{(Lo)}$, $\xi_{(Small)}$ and $\Psi_{(Lo)}$ coefficients in this calculation, because the contract we are calculating for is \emph{short-term} and \emph{low-skill} and the firm size is \emph{small}. 

Determining payoff coefficients ($\Pi_F,\Pi_R,\Pi_T,\Pi_P$) involves the summation of respective \emph{flexibility}, \emph{reliability}, \emph{talent retention} and \emph{potential alternatives} weights in the three constituent states that comprise the GameState. We generate contract distributions and payoff coefficients for the eight GameState with a Jupyter Notebook script.

\vspace{2mm} 

\begin{table}[h!] 
\centering
\begin{tabular}{lrrrr}
\toprule
${}_{GameState}\mkern-6mu\setminus\mkern-6mu{}^{Contract}$ &  Short Low Skill &  Short High Skill &  Long Low Skill &  Long High Skill \\
\midrule
Small Low Bear  &      970.0 &        72.0 &      32.0 &        2.0 \\
Small Low Bull  &      832.0 &        62.0 &      42.0 &        3.0 \\
Large Low Bear  &    27733.0 &      2080.0 &    3200.0 &      240.0 \\
Large Low Bull  &    20800.0 &      1560.0 &    3733.0 &      280.0 \\
Small High Bear &      242.0 &       291.0 &       8.0 &        9.0 \\
Small High Bull &      208.0 &       249.0 &      10.0 &       12.0 \\
Large High Bear &     6933.0 &      8320.0 &     800.0 &      960.0 \\
Large High Bull &     5200.0 &      6240.0 &     933.0 &     1120.0 \\
\bottomrule
\end{tabular}
\caption{GameState Contract Demand Distribution}
\label{table:}
\end{table}

\begin{table}[h!]
\centering

\begin{tabular}{lrrrr}
\toprule
${}_{GameState}\mkern-6mu\setminus\mkern-6mu{}^{Coefficient}$ &  Flexibility &  Reliability &  Talent Retention &  Potential Alternatives \\
\midrule
Small Low Bear  &            4 &            7 &                 0 &                      0 \\
Small Low Bull  &            3 &            2 &                 7 &                      5 \\
Large Low Bear  &            1 &            5 &                 0 &                      0 \\
Large Low Bull  &            0 &            0 &                 2 &                      5 \\
Small High Bear &           11 &           17 &                 3 &                      5 \\
Small High Bull &           10 &           12 &                10 &                     10 \\
Large High Bear &            6 &           12 &                 3 &                      5 \\
Large High Bull &            5 &            7 &                 5 &                     10 \\
\bottomrule
\end{tabular}

\label{Table:}
\caption{GameState Payoff Coefficients}
\end{table}

\newpage

In this section, we introduce the structure of payoff matrices and methods for payoff generation. Each payoff bi-matrix models the payoff for firm and laborer strategy pairs. We generate 5 payoff matrices for each GameState. The first 4 matrices model the 4 contract types in the GameState setting. The fifth matrix incorporates the GameState's contract demand distribution and models the comprehensive GameState with the weighted summation of respective contract payoffs; in other words, we represent the firm payoff with an aggregate of contract payoffs.  

In this first 4x4 matrix, we see that the laborer has 4 strategies; the laborer can participate as a low or high skill gig worker, or a low or high skill employee when competing for a single contract. Similarly, the firm has 4 complementary strategies when hiring for each contract. The highlighted cells in the matrix represent matching strategies. The un-highlighted cells denote a strategy mismatch. In a matched strategy pair, the worker is hired. Conversely, a mismatched strategy pair indicates that no worker is hired, but utility is expended to execute the strategy; mismatched strategy pairs are assigned a marginal negative payoff to reflect this expended utility. Column headers represent firm strategies and row headers represent laborer strategies. Subscript \textit{f} denotes the firm payoff and subscript \textit{l} denotes the laborer payoff.

\vspace{10mm}

\begin{table}[h!] 
\centering
\begin{tabular}{lllll}
\centering
${}_{Laborer}\mkern-6mu\setminus\mkern-6mu{}^{Firm}$ & Gig Lowskill & Gig Highskill & Employee Lowskill & Employee Highskill 
\tabularnewline 
\midrule
Gig Lowskill       &        \cellcolor{blue!25}[$a_{l}$, $a_{f}$] &       [$e_{l}$, $e_{f}$] &            [$i_{l}$, $i_{f}$] &             [$m_{l}$, $m_{f}$] \tabularnewline
Gig Highskill      &        [$b_{l}$, $b_{f}$] &       \cellcolor{blue!25}[$f_{l}$, $f_{f}$] &            [$j_{l}$, $j_{f}$] &             [$n_{l}$, $n_{f}$] \tabularnewline
Employee Lowskill  &        [$c_{l}$, $c_{f}$] &       [$g_{l}$, $g_{f}$] &           \cellcolor{blue!25}[$k_{l}$, $k_{f}$] &             [$o_{l}$, $o_{f}$] \tabularnewline
Employee Highskill &        [$d_{l}$, $d_{f}$] &       [$h_{l}$, $h_{f}$] &            [$l_{l}$, $l_{f}$] &             \cellcolor{blue!25}[$p_{l}$, $p_{f}$] \ 
\label{uncompressedPayoff}
\end{tabular}
\end{table}

\vspace{5mm}

To simplify the game, we first compress the matrix into a 4x2 matrix and thereby eliminate a handful of the mismatched strategies. Later, we will compress the matrix into a 2x2 matrix to further reduce the dimensions of the evolutionary model.

\vspace{5mm}

\noindent\makebox[\textwidth]{
\begin{tabular}{lll}
\centering
${}_{Laborer}\mkern-6mu\setminus\mkern-6mu{}^{Firm}$ &     Gig & Employee \\
Gig Lowskill       &  \cellcolor{blue!25}[$a_{l}$, $a_{f}$ ] &   [$e_{l}$, $e_{f}$ ] \\
Gig Highskill      &  \cellcolor{blue!25}[$b_{l}$, $b_{f}$ ] &   [$f_{l}$, $f_{f}$ ] \\
Employee Lowskill  &  [$c_{l}$, $c_{f}$ ] &   \cellcolor{blue!25}[$g_{l}$ , $g_{f}$ ] \\
Employee Highskill &  [$d_{l}$, $d_{f}$ ] &   \cellcolor{blue!25}[$h_{l}$ ,$h_{f}$ ] \\
\end{tabular}
}

\vspace{5mm}

Only high skill laborers can participate in high-skill contracts. Accordingly, cells $[a_{l}, a_{f}]$ and $[g_{l}, g_{f}]$ become a mismatch for two of our four contract categories: short and long high-skill contracts. In the matrices for these two contract types, the only matching strategies are cells $[b_{l}, b_{f}]$ and $[h_{l}, h_{f}]$ . 

\vspace{5mm}

\noindent\makebox[\textwidth]{
\begin{tabular}{lll}
\toprule
${}_{Laborer}\mkern-6mu\setminus\mkern-6mu{}^{Firm}$ &     Gig & Employee \\
\midrule
Gig Lowskill       &  [$a_{l}$, $a_{f}$ ] &   [$e_{l}$, $e_{f}$ ] \\
Gig Highskill      &  \cellcolor{blue!25}[$b_{l}$, $b_{f}$ ] &   [$f_{l}$, $f_{f}$ ] \\
Employee Lowskill  &  [$c_{l}$, $c_{f}$ ] &   [$g_{l}$ , $g_{f}$ ] \\
Employee Highskill &  [$d_{l}$, $d_{f}$ ] &   \cellcolor{blue!25}[$h_{l}$ , $h_{f}$ ] \\
\bottomrule
\end{tabular}
}

\vspace{5mm}

\subsection{ Firm Strategy Payoffs }
\subsubsection{ Matching Strategies}

To calculate the firm payoff for matching strategies, we apply a general equation that incorporates our GameState assumptions and payoff coefficients.


\begin{equation}
 Payoff_{*} = \frac{\phi(1 - \frac{\beta}{\lambda} - \frac{\Pi_R}{\delta_R+\Theta}  - \frac{\Pi_T}{\delta_T}+ \frac{\Pi_F}{\delta_F})}{\chi}
\end{equation}


\vspace{3mm}

\emph{Compensation} for high skill and low skill contracts respectively costs $100K$ and $30K$ annually regardless of worker type (gig or employee). \emph{Operational Revenue} $\phi$, signifying revenue generated from labor, is 8x the compensation for high skill work and 4x for low skill work; we refer to this factor as the \emph{Revenue Multiplier} $\lambda$. We assume employees receive 40 percent additional compensation in the form of benefits and bonuses; accordingly, the \emph{Benefits Multiplier} $\beta$ is 1.4 for employee and 1 for gig strategies. The first component of firm payoff debits the bonus-adjusted-compensation from operational revenue. GameState payoff coefficients, denoted with $\Pi$, incorporate additional hiring considerations that impact a strategy's overall payoff. Discount values, denoted with $\delta$, are applied to each payoff coefficient to account for gig worker and employee disparities. Larger discount values further reduce the importance of the payoff coefficient. Subscripts R, F and T are used to indicate discounts and coefficients for \emph{reliability}, \emph{flexibility} and \emph{talent retention} respectively. \emph{Reliability} $\Pi_R$ (the importance of reliability expressed through the cost of expected error) and \emph{talent retention} $\Pi_T$ (the cost of labor acquisition and retention) detract from the payoff while labor \emph{flexibility} $\Pi_F$ is an asset. Hiring considerations denoted by payoff coefficients directly impact operational revenue. Finally, we standardize the annual payoff to measure for the interval of the contract, 2 or 26 weeks depending on the contract length; $\chi$ equates to 52 weeks divided by the contract duration in weeks.

 Gig and employee strategies are assigned a $\delta_R$ of 8 and 40 respectively, indicating that employees are 7.5x more reliable than gig workers. Further, talent retention is more expensive for employees than for gig workers. We assign gig and employee strategies a $\delta_T$ of 200 and 50 respectively, indicating that employees are 4x more expensive for firms to acquire and retain. Since gig arrangements enable on-demand labor, gig workers provide more flexibility to the firm. Therefore, gig and employee strategies are assigned a discount $\delta_F$ of 8 and 60 respectively, indicating gig workers provide 6.25x more labor flexibility than employees. We introduce an \emph{OverSkill} $\Theta$ discount of +5 if a high skill worker engages in a low skill contract; this overskill mismatch implies a reduced cost of expected error when a high skill worker takes on low skill tasks. Below, we break down the specific firm payoffs for each potential matching strategy. 
\\\\
\noindent\makebox[\textwidth]{
\begin{tabular}{lll}
\toprule
${}_{Laborer}\mkern-6mu\setminus\mkern-6mu{}^{Firm}$&     Gig & Employee \\
\midrule
Gig Lowskill       &  [$a_{l}$, \colorbox{blue!25}{$a_{f}$}] &   [$e_{l}$, $e_{f}$ ] \\
Gig Highskill      &  [$b_{l}$, \colorbox{blue!25}{$b_{f}$}] &   [$f_{l}$, $f_{f}$ ] \\
Employee Lowskill  &  [$c_{l}$, $c_{f}$ ] &   [$g_{l}$, \colorbox{blue!25}{$g_{f}$}] \\
Employee Highskill &  [$d_{l}$, $d_{f}$] &   [$h_{l}$, \colorbox{blue!25}{$h_{f}$}] \\
\bottomrule
\end{tabular}
}
\vspace{3mm}

\begin{equation}
\centering
    \colorbox{blue!25}{$a_{f}$} =  \frac{\phi(1 - \frac{1}{\lambda} - \frac{\Pi_R}{8}  - \frac{\Pi_T}{200}+ \frac{\Pi_F}{8})}{\chi}
\end{equation}

\begin{equation}
\centering
    \colorbox{blue!25}{$b_{f}$} =  \frac{\phi(1 - \frac{1}{\lambda} - \frac{\Pi_R}{8+\Theta}  - \frac{\Pi_T}{200}+ \frac{\Pi_F}{8})}{\chi}
\end{equation}

\begin{equation}
\centering
    \colorbox{blue!25}{$g_{f}$}= \frac{\phi(1 - \frac{1.4}{\lambda} - \frac{\Pi_R}{40}  - \frac{\Pi_T}{50}+ \frac{\Pi_F}{60})}{\chi}
\end{equation}

\begin{equation}
\centering
    \colorbox{blue!25}{$h_{f}$}=  \frac{\phi(1 - \frac{1.4}{\lambda} - \frac{\Pi_R}{40+\Theta}  - \frac{\Pi_T}{50}+ \frac{\Pi_F}{60})}{\chi}
\end{equation}
\vspace{3mm}
\subsubsection{ Mismatched Strategies}

For mismatching strategies, we assign a negative payoff to reflect the firm's utility expenditure in pursuing the strategy of hiring a category of worker but failing to hire. The assigned payoff is the negative value of one fiftieth of the contract's operational revenue over the contract duration. As low skill laborers can not work high skill contracts, $a_{l}$ and $g_{l}$ become mismatching strategies and assigned the mismatched strategy payoff. For low skill contracts, strategies highlighted in pink are mismatching strategies. For high skill contracts, strategies highlighted in pink and yellow are mismatching strategies.

\noindent\makebox[\textwidth]{
\begin{tabular}{lll}
\toprule
${}_{Laborer}\mkern-6mu\setminus\mkern-6mu{}^{Firm}$ &     Gig & Employee \\
\midrule
Gig Lowskill       &  [$a_{l}$, \colorbox{yellow}{$a_{f}$} ] &   [ $e_{l}$, \colorbox{pink}{$e_{f}$}] \\
Gig Highskill      &  [$b_{l}$, $b_{f}$ ] &   [ $f_{l}$, \colorbox{pink}{$f_{f}$} ] \\
Employee Lowskill  &  [$c_{l}$, \colorbox{pink}{$c_{f}$} ] &   [$g_{l}$ , \colorbox{yellow}{$g_{f}$} ] \\
Employee Highskill &  [ $d_{l}$, \colorbox{pink}{$d_{f}$} ] &   [$h_{l}$ , $h_{f}$ ] \\
\bottomrule
\end{tabular}
}

\vspace{5mm}

\begin{equation}
\centering
    \colorbox{pink}{$e_{f},f_{f},c_{f}, d_{f},$}\colorbox{yellow}{$ a_{f}, g_{f}$}= -\frac{\phi}{50\chi}
\end{equation}

\subsection{ Laborer Strategy Payoffs }

\subsubsection{ Matching Strategies}

To calculate the laborer payoffs for matching strategies, we apply the following general equation. 

\begin{equation}
    Payoff_{*} = \frac{\Psi(\beta\epsilon + \frac{\Pi_P}{\delta_P})}{\chi}
\end{equation}

\emph{Compensation} $\Psi$ for high skill and low skill contracts respectively costs $100K$ and $30K$ annually regardless of worker type (gig or employee). In a bull market, we assume that workers experience a 5 percent rate of involuntary attrition, translating into an \emph{employment stability} coefficient, denoted with $\epsilon$, of 0.95.  In a bear market, we assign a value of 0.7 to $\epsilon$, implying a 30 percent rate of involuntary attrition. In addition to receiving compensation for labor, laborers receive a payoff from potential alternative engagements outside of their primary contracts. $\Pi_P$ and $\delta_P$ denote the potential alternatives payoff coefficient and discount respectively. As gig arrangements champion flexibility and self governance, we assume that gig workers have the opportunity to take part in 5x the potential alternative engagements compared to employees. In our model, this translates to a $\delta_P$ of 5 and 25 for gig and employee strategies respectively.

We adjust parameters in specific scenarios to account for additional phenomena. Employees competing for high skill contract during a bull market will receive 3x additional benefits, an optimistic gratuity, to account for compounding stock options or carried interest bonus. During a bear market, low skill workers, especially those with short term work arrangements, are more adversely affected as industries discharge commodity skill laborers. We subtract 8 from $\Pi_P$ if a low skill worker competes for a gig contract during a bear market. As unemployment rates hike during a bear market, low skill firms begin to hire high skill workers. Accordingly, high skill workers gain access to a broader set of alternative work options during a bear market. Since a gig worker has increased flexibility, high skill gig workers benefit the most from this increased opportunity. To account for this, we add 15 to $\Pi_P$ for high skill gig strategy payoffs during a bear market.

\vspace{3mm}
\noindent\makebox[\textwidth]{
\begin{tabular}{lll}
\toprule
${}_{Laborer}\mkern-6mu\setminus\mkern-6mu{}^{Firm}$&     Gig & Employee \\
\midrule
Gig Lowskill       &  [\colorbox{blue!25}{$a_{l}$}, $a_{f}$ ] &   [$e_{l}$, $e_{f}$ ] \\
Gig Highskill      &  [\colorbox{blue!25}{$b_{l}$}, $b_{f}$ ] &   [$f_{l}$, $f_{f}$ ] \\
Employee Lowskill  &  [$c_{l}$, $c_{f}$ ] &   [\colorbox{blue!25}{$g_{l}$} , $g_{f}$ ] \\
Employee Highskill &  [$d_{l}$, $d_{f}$ ] &   [\colorbox{blue!25}{$h_{l}$} ,$h_{f}$ ] \\
\bottomrule
\end{tabular}
}

\vspace{5mm}

\begin{equation}
\centering
    \colorbox{blue!25}{$a_{l}$} = \frac{\Psi(\epsilon + \frac{\Pi_P}{5})}{\chi}
\end{equation}

\begin{equation}
\centering
    \colorbox{blue!25}{$b_{l}$} = \frac{\Psi(\epsilon + \frac{\Pi_P}{5})}{\chi}
\end{equation}

\begin{equation}
\centering
    \colorbox{blue!25}{$g_{l}$} = \frac{\Psi(1.4\epsilon + \frac{\Pi_P}{25})}{\chi}
\end{equation}

\begin{equation}
\centering
    \colorbox{blue!25}{$h_{l}$} = \frac{\Psi(1.4\epsilon + \frac{\Pi_P}{25})}{\chi}
\end{equation}

\subsubsection{Mismatched Strategies}

For mismatching strategies, we assign a negative payoff to reflect the laborer's utility expenditure in pursuing an employment strategy in competition of a contract but failing to get hired. The assigned payoff is the negative value of one fiftieth of the contract's compensation over the contract duration. As low skill laborers can not work high skill contracts, $a_{l}$ and $g_{l}$ become mismatching strategies and assigned the mismatched strategy payoff. For low skill contracts, strategies highlighted in pink are mismatched strategies. For high skill contracts, strategies highlighted in pink and yellow are mismatched strategies.

\noindent\makebox[\textwidth]{
\begin{tabular}{lll}
\toprule
${}_{Laborer}\mkern-6mu\setminus\mkern-6mu{}^{Firm}$&     Gig & Employee \\
\midrule
Gig Lowskill       &  [\colorbox{yellow}{$a_{l}$}, $a_{f}$ ] &   [\colorbox{pink}{$e_{l}$}, $e_{f}$ ] \\
Gig Highskill      &  [$b_{l}$, $b_{f}$ ] &   [\colorbox{pink}{$f_{l}$}, $f_{f}$ ] \\
Employee Lowskill  &  [\colorbox{pink}{$c_{l}$}, $c_{f}$ ] &   [\colorbox{yellow}{$g_{l}$} , $g_{f}$ ] \\
Employee Highskill &  [\colorbox{pink}{$d_{l}$}, $d_{f}$ ] &   [$h_{l}$ , $h_{f}$ ] \\
\bottomrule
\end{tabular}
}
\vspace{5mm}
\begin{equation}
\centering
    \colorbox{pink}{$e_{l},f_{l},c_{l}, d_{l},$}\colorbox{yellow}{$ a_{l}, g_{l}$}= - \frac{\Psi}{50\chi}
\end{equation}
\vspace{3mm}
\subsection{GameState Aggregate: Weighted Payoff}

The fifth matrix incorporates the GameState's  contract distribution and models the payoff for the GameState; while the first four matrices represent payoffs for each contract, the fifth matrix represents the payoff for an entire firm in a market setting. Since the firm consists of a distribution of contracts, we calculate the GameState payoff by taking the weighted summation of contract payoffs. Below, GS denotes the GameState payoff. $\alpha$ represents the firm demand for each contract type in a single GameState. We previously calculated these values when generating our GameState contract demand distributions. Subscript 1,2,3 and 4 respectively denote short low skill, short high skill, long low skill and long high skill contracts. 

\noindent 

\vspace{5mm}

\begin{equation}
GS = \alpha^{}_{1}
\begin{pmatrix}
 [a_{l}, a_{f} ] &   [e_{f},e_{l}] \\
 [b_{l}, b_{f}] &   [ f_{l}, f_{f} ] \\
 [c_{l}, c_{f} ] &   [g_{l} , g_{f} ] \\
 [d_{l}, d_{f}] &   [h_{l} , h_{f} ] \\
\end{pmatrix}_{1}
+
\alpha^{}_{2}
\begin{pmatrix}
 [a_{l}, a_{f} ] &   [e_{f},e_{l}] \\
 [b_{l}, b_{f}] &   [ f_{l}, f_{f} ] \\
 [c_{l}, c_{f} ] &   [g_{l} , g_{f} ] \\
 [d_{l}, d_{f}] &   [h_{l} , h_{f} ] \\
\end{pmatrix}_{2}
+
\alpha^{}_{3}
\begin{pmatrix}
 [a_{l}, a_{f} ] &   [e_{f},e_{l}] \\
 [b_{l}, b_{f}] &   [ f_{l}, f_{f} ] \\
 [c_{l}, c_{f} ] &   [g_{l} , g_{f} ] \\
 [d_{l}, d_{f}] &   [h_{l} , h_{f} ] \\
\end{pmatrix}_{3}
+
\alpha^{}_{4}
\begin{pmatrix}
 [a_{l}, a_{f} ] &   [e_{f},e_{l}] \\
 [b_{l}, b_{f}] &   [ f_{l}, f_{f} ] \\
 [c_{l}, c_{f} ] &   [g_{l} , g_{f} ] \\
 [d_{l}, d_{f}] &   [h_{l} , h_{f} ] \\
\end{pmatrix}_{4}
\end{equation}
\vspace{5mm}

\subsection{Payoff Matrices}

We generate payoff matrices for the eight GameStates with a Jupyter Notebook script that implements all of our assumptions.

We reduce GameState payoff matrices from 4x2 to 2x2 bi-matrices to further truncate model dimensions. We combine high and low skill strategy payoffs for gig and employee strategies respectively. Therefore in the 2x2 bi-matrix, the 2x2 gig strategy accounts for high and low skill gig strategies and the 2x2 employee strategy accounts for high and low skill employee strategies. We implement this matrix reduction process with a Jupyter Notebook script and generate 8 GameState payoff matrices, which can be found in Appendix section C.1-8. 

\vspace{5mm}
\noindent\makebox[\textwidth]{
\begin{tabular}{lll}
\centering
${}_{Laborer}\mkern-6mu\setminus\mkern-6mu{}^{Firm}$&     Gig & Employee \\
Gig Lowskill       &  [$a_{l}$, $a_{f}$ ] &   [$e_{l}$, $e_{f}$ ] \\
Gig Highskill      &  [$b_{l}$, $b_{f}$ ] &   [$f_{l}$, $f_{f}$ ] \\
Employee Lowskill  &  [$c_{l}$, $c_{f}$ ] &   [$g_{l}$ , $g_{f}$ ] \\
Employee Highskill &  [$d_{l}$, $d_{f}$ ] &   [$h_{l}$ , $h_{f}$ ] \\
\end{tabular}

\begin{tabular}{lll}
\centering
${}_{Laborer}\mkern-6mu\setminus\mkern-6mu{}^{Firm}$&     Gig & Employee \\
Gig  &  [$a_{l}+b_{l}$,$a_{f}+b_{f}$ ] &   [$e_{l}+f_{l}$, $e_{f}+f_{f}$]  \\
Employee &  [$c_{l}+d_{l}$, $c_{f}+d_{f}$ ] &   [$g_{l}+h_{l}$, $g_{f}+h_{f}$ ] \\
\end{tabular}
}
\\\\

\section{Evolutionary Model}

\subsection{System Equilibria}
In this section, we solve for our evolutionary system's fixed points for the general case. For each fixed point, we analyze the stability of the equilibrium and offer an explanation. 

\subsubsection{Fixed Points}
Solving our system of two equations and two unknowns, we reach a general solution set that contains five fixed points. Below, we list each fixed point in the form $(x_1, y_1)^*$.\\\\
$FixedPoint_1=(0,0)^*$\\\\
$FixedPoint_2=(1,0)^*$\\\\
$FixedPoint_3=(0,1)^*$\\\\
$FixedPoint_4=(1,1)^*$\\\\
$FixedPoint_5= (\frac{-(c_{f0} - d_{f0} - c_{f0}n + c_{f1}n + d_{f0}n - d_{f1}n)}{(a_{f0} - b_{f0} - c_{f0} + d_{f0} - a_{f0}n + a_{f1}n + b_{f0}n - b_{f1}n + c_{f0}n - c_{f1}n - d_{f0}n + d_{f1}n)}$, \\\\ 
\hspace*{26mm}$\frac{-(b_{l0} - d_{l0} - b_{l0}n + b_{l1}n + d_{l0}n - d_{l1}n)}{(a_{l0} - b_{l0} - c_{l0} + d_{l0} - a_{l0}n + a_{l1}n + b_{l0}n - b_{l1}n + c_{l0}n - c_{l1}n - d_{l0}n + d_{l1}n)})^*
$
\\

\noindent FixedPoints 1,2,3 and 4 lie on the extremes of our system, and $FixedPoint_5$ is our only internal equilibrium. In our model, this implies that $FixedPoint_5$ is the only equilibrium with a co-existence of gig workers and employees.

\subsubsection{Stability Analysis}
To analyze the stability of each equilibrium, we examine the eigenvalues of the Jacobian matrix for each fixed point. For fixed point to be asymptotically stable, eigenvalues of the Jacobian must have all negative real parts. If eigenvalues have all positive real parts, the fixed point is unstable. If the set of eigenvalues includes both positive and negative real parts, the equilibrium is a saddle point.

In order to feasibly analyze the negativity of eigenvalues, we reduce the number of generalized parameters. Since mismatching strategy payoffs are assigned marginal values in respect to matching strategy payoffs, we set them to 0. To further simplify, we demonstrate our stability analysis with $n=0$ rather than allowing n to remain a generalized parameter. The remaining parameters are matching strategy payoffs; for all GameStates, these payoffs take positive values. Given these assumptions, the simplified Jacobian is as follows. 

\begin{equation}
J = \begin{bmatrix}
-(2x_1 - 1)(a_{l0}y_1 - d_{l0} + d_{l0}y_1) & -x_1(a_{l0} + d_{l0})(x_1 - 1)\\
-y_1(a_{f0} + d_{f0})(y_1 - 1) & -(2y_1 - 1)(a_{f0}x_1 - d_{f0} + d_{f0}x_1)
\end{bmatrix}_{(x_1^*, y_1^*)}
\end{equation}

\noindent \textbf{Saddle Points}\\\\
We find that our internal equilibrium $FixedPoint_5$ is a saddle point. The set of eigenvalues always takes both positive and negative values as the two eigenvalues are opposites of each other. 
\\\\
 \indent $Eigenvalue_1=\frac{(a_{f0}a_{l0}d_{f0}d_{l0}(a_{f0} + d_{f0})(a_{l0} + d_{l0}))^{(1/2)}}{(a_{f0}a_{l0} + a_{f0}d_{l0} + a{l0}d_{f0} + d_{f0}d_{l0})}$              \\\\                      \indent $Eigenvalue_2=-\frac{(a_{f0}a_{l0}d_{f0}d_{l0}(a_{f0} + d_{f0})(a_{l0} + d_{l0}))^{(1/2)}}{(a_{f0}a_{l0} + a_{f0}d_{l0} + a{l0}d_{f0} + d_{f0}d_{l0})}$
 \\

\noindent \textbf{Unstable Fixed Points}
\\\\
We find that $FixedPoint_2$ $(1,0)^*$ and $FixedPoint_3$ $(0,1)^*$, equilibria at mismatching extremes, are unstable. Since all matching strategy payoffs are positive, both eigenvalues of the Jacobian matrix for each of the two fixed points are always positive.  \\

For $FixedPoint_2$, $ Eigenvalue_1=a_{f0}$ and
$Eigenvalue_2=d_{l0}$.\\

For $FixedPoint_3$, $ Eigenvalue_1=a_{l0}$ and
$Eigenvalue_2=d_{f0}$.\\

\noindent If the system begins at one of these unstable fixed points, the system will not remain stationary. Rather, the system will evolve on a trajectory towards a stable fixed point.
\\\\
\noindent \textbf{Stable Fixed Points}
\\\\
We find that $FixedPoint_1$ $(0,0)^*$ and $FixedPoint_4$ $(1,1)^*$, equilibria at matching extremes, are unstable. Since all matching strategy payoffs are positive, both eigenvalues of the Jacobian matrix for each of the two fixed points are always negative.\\

For $FixedPoint_1$, $ Eigenvalue_1=-d_{f0}$ and
$Eigenvalue_2=-d_{l0}$.\\

For $FixedPoint_4$, $ Eigenvalue_1=-a_{f0}$ and
$Eigenvalue_2=-a_{l0}$.\\

\noindent If either the initial condition begins at or the system evolves to one of these stable fixed points, the system will remain stationary. These two stable fixed points are our Evolutionary Stable Strategies (ESS). At $(0,0)^*$, firms and laborers both have a density of 0 for the gig strategy, implying that both populations consist entirely of employee strategies. At $(1,1)^*$, firm and laborer populations are fully dominated by gig strategies. If the system evolves to an ESS, no auxiliary strategies will be able to invade the dominating strategy population given an initially low strategy density \citep{taylor1978evolutionary, smith1973logic}. In other words, if the labor market evolves to a stage where both laborers and firms consist entirely of gig strategies, the system will forever remain fixed, implying that gig workers will dominate the labor market forever and that there will never exist an employee strategy again. Likewise, if the labor market evolves to a stage where both laborers and firms are comprised entirely of employee strategies, the system will remain fixed and employee strategies will dominate the labor market forever. While most studies in evolutionary game theory focus on the evolutionary outcome to an ESS, in this work, we instead propose that the system will never evolve to an ESS; we expand on this notion in succeeding sections.

\subsection{Oscillating Replicator Dynamics}
\subsubsection{Computational Notes}

We generate our evolutionary diagrams with Matlab and our phase diagrams with Mathematica. We also employ Matlab for calculating attractor arc reference points, fixed points, the Jacobian, eigenvalues and streamplot equations. We use the Adobe Photoshop editor for superimposing diagrams and incorporating additional visual aids. 

\subsubsection{Trapping Zone Orbit}

We select our initial condition to be $(0.45,0.40)$, the attractor position at $n=0.5$, an approximation for a point in the trapping zone. This selection implies that we assume our system has previously oscillated in the trapping zone up until this moment in time. This assumption is sensible because the labor market maintains a co-existence of gig workers and employees. 
\begin{figure*}[h!]
    \centering
    \makebox[\textwidth]{
    \centering
    \includegraphics[width=17cm]{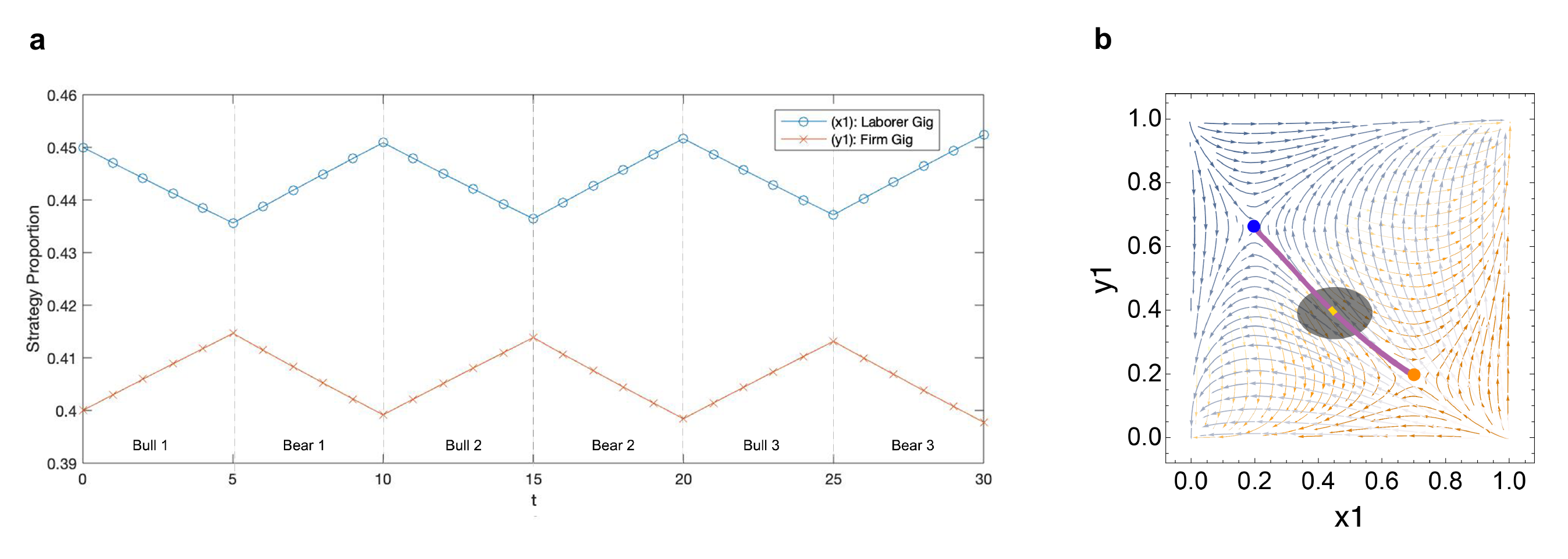}
    }
    \caption{Trapping Zone Oscillation with Initial Conditions $(0.45, 0.40)$, $\omega = 0.005$ and $n=1$ and Theoretical GameState Pair, see Figure 1. (a) Mismatching Oscillatory Behavior in Trapping Zone (b) Trapping Zone Orbit. We illustrate the trapping zone orbit in yellow. A reference attractor arc is plotted in purple and attractor positions at $n=0$ and $n=1$ are represented in orange and blue respectively. The opaque black ellipse is a background element to help visually contrast with the trapping zone. }
\end{figure*}
\subsubsection{Escape Demonstration with Different Initial Conditions}
In this section, we illustrate an example of \emph{escape} by increasing $\omega$ by a factor of 20, $\omega=0.1$, such that the system reaches escape boundary. In Figure 21, the increase in $\omega$ occurs at the start of the bull market. In Figure 22, the increase in $\omega$ occurs at the start of the bear market. The purpose of the following demonstration is to illustrate how initial conditions can alter escape destination. Therefore, claims founded on escape destination are indefensible because escape destination is determined by arbitrary preparations of initial conditions.
\begin{figure*}[h!]
    \centering

\noindent\makebox[\textwidth]{
\centering

\includegraphics[width=15cm]{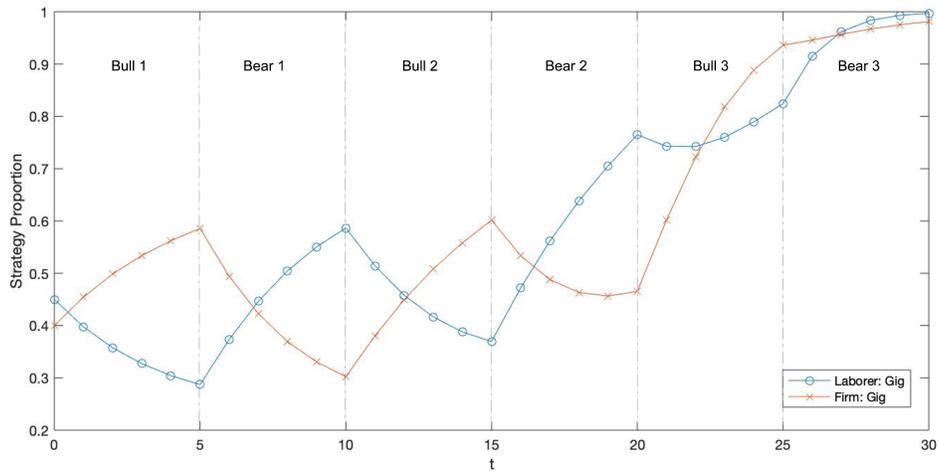}
}
   \caption{Escape Demonstration with Initial Conditions $(0.45, 0.40)$, $\omega = 0.1$ and $n=1$ and Theoretical GameState Pair, see Figure 1.}
\end{figure*}

\begin{figure*}[h!]
\noindent\makebox[\textwidth]{
\centering
\includegraphics[width=15cm]{figure_22.pdf}
}
 \caption {Escape Demonstration with Initial Conditions $(0.45, 0.40)$, $\omega = 0.1$ and $n=0$
 and Theoretical GameState Pair, see Figure 1.}
\end{figure*}

\newpage
\section{Payoff Matrices}
\subsection{ Small Low Bear }
\begin{table}[h!] 
\centering
\begin{tabular}{lll}
\toprule
${}_{Laborer}\mkern-6mu\setminus\mkern-6mu{}^{Firm}$ &                     Gig &                Employee \\
\midrule
Gig      &  [1066048.0, 8397092.0] &   [-78764.0, -375840.0] \\
Employee &   [-78764.0, -375840.0] &  [3494832.0, 9154366.0] \\
\bottomrule
\end{tabular}

\caption{2x2 Payoff Matrix for GameState in Setting Small Low Bear }
\label{table:}
\end{table}
\vspace{-3mm}

\newpage

\subsection{ Small Low Bull }

\begin{table}[h!] 
\centering
\begin{tabular}{lll}
\toprule
${}_{Laborer}\mkern-6mu\setminus\mkern-6mu{}^{Firm}$ &                      Gig &                Employee \\
\midrule
Gig      &  [6950788.0, 14231744.0] &   [-78896.0, -378148.0] \\
Employee &    [-78896.0, -378148.0] &  [6558500.0, 8587992.0] \\
\bottomrule
\end{tabular}

\caption{2x2 Payoff Matrix for GameState in Setting Small Low Bull }
\label{table:}
\end{table}
\vspace{-3mm}

\subsection{ Large Low Bear }
\begin{table}[h!] 
\centering
\begin{tabular}{lll}
\toprule
${}_{Laborer}\mkern-6mu\setminus\mkern-6mu{}^{Firm}$ &                        Gig &                    Employee \\
\midrule
Gig      &  [57566264.0, 293689368.0] &   [-3991878.0, -19181272.0] \\
Employee &  [-3991878.0, -19181272.0] &  [175951620.0, 462572232.0] \\
\bottomrule
\end{tabular}

\caption{2x2 Payoff Matrix for GameState in Setting Large Low Bear }
\label{table:}
\end{table}
\vspace{  -3mm}

\newpage

\subsection{ Large Low Bull }
\begin{table}[h!] 
\centering
\begin{tabular}{lll}
\toprule
${}_{Laborer}\mkern-6mu\setminus\mkern-6mu{}^{Firm}$ &                         Gig &                    Employee \\
\midrule
Gig      &  [350581940.0, 608754400.0] &   [-3993720.0, -19185200.0] \\
Employee &   [-3993720.0, -19185200.0] &  [331969900.0, 512758880.0] \\
\bottomrule
\end{tabular}

\caption{2x2 Payoff Matrix for GameState in Setting Large Low Bull }
\label{table:}
\end{table}
\vspace{  -3mm}

\subsection{ Small High Bear }
\begin{table}[h!] 
\centering
\begin{tabular}{lll}
\toprule
${}_{Laborer}\mkern-6mu\setminus\mkern-6mu{}^{Firm}$ &                     Gig &                Employee \\
\midrule
Gig      &  [8062392.0, 2386627.0] &   [-78164.0, -565658.0] \\
Employee &   [-78164.0, -565658.0] &  [2762366.0, 7506322.0] \\
\bottomrule
\end{tabular}

\caption{2x2 Payoff Matrix for GameState in Setting Small High Bear }
\label{table:}
\end{table}
\vspace{  -3mm}

\subsection{ Small High Bull }
\begin{table}[h!] 
\centering
\begin{tabular}{lll}
\toprule
${}_{Laborer}\mkern-6mu\setminus\mkern-6mu{}^{Firm}$ &                     Gig &                Employee \\
\midrule
Gig      &  [6864878.0, 9219843.0] &   [-77416.0, -560542.0] \\
Employee &   [-77416.0, -560542.0] &  [8452547.0, 6917537.0] \\
\bottomrule
\end{tabular}

\caption{2x2 Payoff Matrix for GameState in Setting Small High Bull }
\label{table:}
\end{table}
\vspace{  -3mm}

\subsection{ Large High Bear }
\begin{table}[h!] 
\centering
\begin{tabular}{lll}
\toprule
${}_{Laborer}\mkern-6mu\setminus\mkern-6mu{}^{Firm}$ &                         Gig &                    Employee \\
\midrule
Gig      &  [410392908.0, 101342039.0] &   [-3983558.0, -28789272.0] \\
Employee &   [-3983558.0, -28789272.0] &  [139995466.0, 413857629.0] \\
\bottomrule
\end{tabular}

\caption{2x2 Payoff Matrix for GameState in Setting Large High Bear }
\label{table:}
\end{table}

\newpage

\subsection{ Large High Bull }
\begin{table}[h!] 
\centering
\begin{tabular}{lll}
\toprule
${}_{Laborer}\mkern-6mu\setminus\mkern-6mu{}^{Firm}$ &                         Gig &                    Employee \\
\midrule
Gig      &  [352366500.0, 474091576.0] &   [-3987480.0, -28791200.0] \\
Employee &   [-3987480.0, -28791200.0] &  [433984460.0, 467401118.0] \\
\bottomrule
\end{tabular}

\caption{2x2 Payoff Matrix for GameState in Setting Large High Bull }
\label{table:}
\end{table}

\subsection{Vertical Attractor Arc}

\begin{table}[h!] 
\begin{tabular}{c}
    \makebox[\textwidth]{
     \includegraphics[width=13cm]{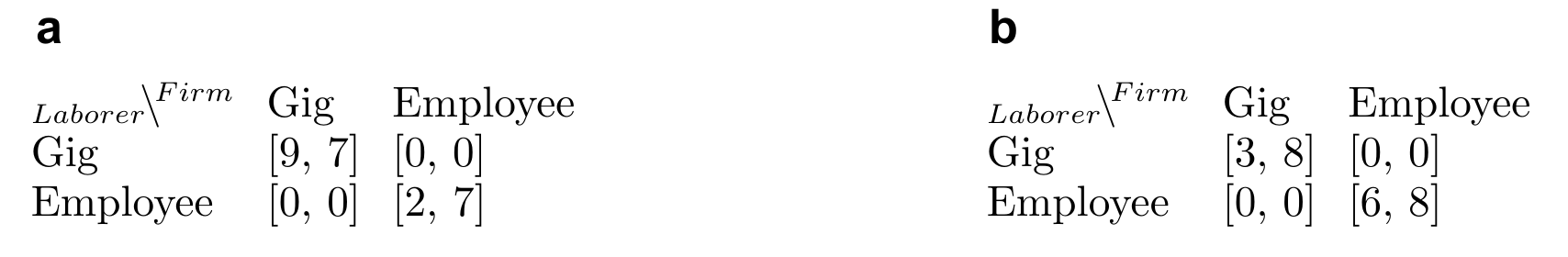}   
     }
     \end{tabular}
    \caption{Payoffs for Vertical Attractor Arc Demonstration. (a) Bear Market, $n=0$ (b) Bull Market, $n=1$}
\end{table}
%

\subsection{Horizontal Attractor Arc }

\begin{table}[h!] 
\begin{tabular}{c}
    \makebox[\textwidth]{
      \includegraphics[width=13cm]{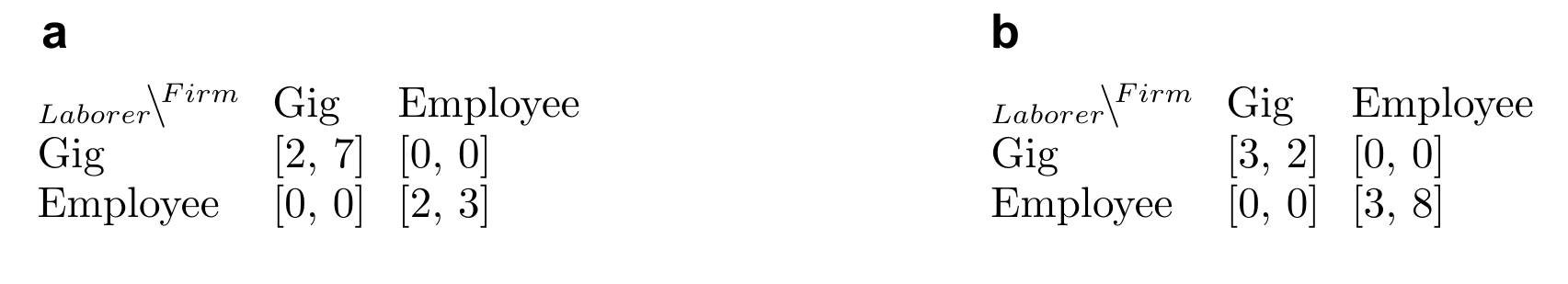}
     }
     \end{tabular}
    \caption{2x2 Payoffs for Horizontal Attractor Arc Demonstration. (a) Bear Market, $n=0$ (b) Bull Market, $n=1$}
\end{table}

\end{document}